# *dciWebMapper2*: Enhancing the *dciWebMapper* framework toward integrated, interactive visualization of linked multi-type maps, charts, and spatial statistics and analysis

Sarigai Sarigai[1,2], Liping Yang[1,2,3]*, Katie Slack[1,2], Carolyn Fish[4], Michaela Buenemann[5], Qiusheng Wu[6], Yan Lin[7], Joseph A. Cook[8,9], David Jacobs[10]

\* Corresponding author: <lipingyang@unm.edu>

[1] Department of Geography and Environmental Studies, University of New Mexico, Albuquerque, NM 87131, USA
[2] Center for the Advancement of Spatial Informatics Research and Education (ASPIRE), University of New Mexico, Albuquerque, NM 87131, USA
[3] Department of Computer Science, University of New Mexico, Albuquerque, NM 87106, USA
[4] Department of Geography, University of Oregon, Eugene, OR 97403, USA
[5] Department of Geography and Environmental Studies, New Mexico State University, Las Cruces, NM 88003, USA
[6] Department of Geography & Sustainability, University of Tennessee, Knoxville, TN 37996, USA
[7] Department of Geography, The Pennsylvania State University, University Park, PA 16802
[8] Department of Biology, University of New Mexico, Albuquerque, NM 87106, USA
[9] Museum of Southwestern Biology, University of New Mexico, Albuquerque, NM 87106, USA
[10] Geospatial and Population Studies, University of New Mexico, Albuquerque, NM 87131, USA

**ABSTRACT**

As interactive web-based geovisualization becomes increasingly vital across disciplines, there is a growing need for open-source frameworks that support dynamic, multi-attribute spatial analysis and accessible design. This paper introduces *dciWebMapper2*, a significant expansion of the original *dciWebMapper* framework, designed to enable exploratory analysis across domains such as climate justice, food access, and social vulnerability. The enhanced framework integrates multiple map types, including choropleth, proportional symbol, small multiples, and heatmaps, with linked statistical charts (e.g., scatter plots, boxplots) and time sliders, all within a coordinated-view environment. Dropdown-based controls allow flexible, high-dimensional comparisons while maintaining visual clarity. Grounded in cartographic and information visualization principles, *dciWebMapper2* is fully open-source, self-contained, and server-free, supporting modularity, reproducibility, and long-term sustainability. Three applied use cases demonstrate its adaptability and potential to democratize interactive web cartography. This work offers a versatile foundation for inclusive spatial storytelling and transparent geospatial analysis in research, education, and civic engagement.

**Keywords**: Interactive web mapping; coordinated views; geovisual analytics; exploratory spatial analysis; linked maps and charts; web GIS application; open-source geospatial tools

# 1. Introduction and motivation

Web-based mapping frameworks have become increasingly important, not only in cartography and GIScience but also across a wide range of domains where the analysis, visualization, and communication of geospatial big data are essential. *These frameworks offer accessible and interactive platforms for exploring spatial information and revealing invaluable insights that might otherwise remain invisible* (Sarigai et al., 2025). As geospatial datasets grow in volume and complexity—often incorporating diverse formats, temporal dimensions, and associated statistical attributes—and as web technologies advance, *there is a critical need for open-source tools built on flexible and extensible platforms supporting multi-dimensional interactive exploration* (Sarigai et al., 2025; MacEachren and Kraak 2001; Roth, 2015; Roth, 2021). In particular, frameworks that integrate diverse map types (e.g., choropleth map and small multiples with a dropdown) and link those diverse map types with interactive chart types (e.g., scatter plots, histograms, boxplots, stacked bar charts) and spatial statistics (e.g., linear regression visuals) enhance how spatial patterns and relationships can be explored, interpreted, and communicated. *These capabilities are especially important in domains where understanding geographic variation across space, time, and attributes is critical*, such as public health, transportation safety, environmental conditions, social justice, food security, and economic factors, *and where users benefit from dynamic, coordinated-view interactions between spatial and non-spatial representations.*

In an effort to address the needs mentioned above, a lightweight web mapping framework, named *dciWebMapper* (Sarigai et al., 2025), was previously developed to support the rapid creation of interactive cartographic applications using current web standards. Importantly, *maps integrate three fundamentally different dimensions of meaning: space, time, and attributes situated in space-time* (MacEachren, 2004). While the original *dciWebMapper* framework emphasized map-centered interaction and user-focused customization, it offered limited support for diverse visualization modes and statistical integration. This paper presents a substantial extension of the *dciWebMapper* framework, introducing integrated capabilities for multi-type map displays (e.g., choropleth maps, small multiples, and temporal visualizations), interactive statistical charts (e.g., boxplots and histograms), and spatial statistical representations (e.g., scatterplots with regression lines, spatial selection/filter via points in polygon, and hotspot detection via heatmaps). In particular, in *dciWebMapper2* spatial meaning is conveyed through a variety of base and thematic map types; temporal meaning is incorporated through interactive time sliders; and the interplay of space and attributes is visualized through maps, interactive charts, filters, and linked data tables. Importantly, these components are crossfilterable, that is, user interaction with any one element (e.g., selecting a time range or filtering by category) dynamically updates the other elements, enabling seamless, bidirectional exploration across spatial, temporal, and attribute dimensions. *The goal of this expansion is not to replace the original dciWebMapper framework, but to build upon and expand it*, enabling more advanced exploratory data analysis (EDA) and more flexible, data-driven cartographic communication. By supporting coordinated-views across spatial (maps) and non-spatial elements (e.g., time slider and charts from attributes), the enhanced *dciWebMapper2* framework advances interactive web cartography and contributes to the broader development of open, extensible, and transferable tools for spatial data exploration, storytelling, analysis, and science communication.

## 2. Background
This section reviews developments in cartography, Geographic Information Science (GIS), geovisualization, and open-source web mapping, emphasizing coordinated-views, interface redundancy, and open science for accessible, interactive visualization.

### 2.1. Cartography, GIS, geovisualization, and interactive web maps

Cartography has evolved from a discipline balancing aesthetics, symbolization, and spatial logic into a dynamic, data-driven practice shaped primarily by advances in GIS and digital technologies, with additional influence from remote sensing (Meynen, 1973; Monmonier, 1996; MacEachren, 2004; Silayo, 2002). The rise of GIS in the 1960s marked a pivotal shift toward computational spatial analysis, enabling the integration of geographic features with structured attribute data to support sophisticated reasoning and decision-making (Department of the Environment, 1987; Olson, 2001; Jones, 2014; Waters, 2018). This synergy between GIS and cartography laid the groundwork for geovisualization, a field that synthesizes spatial analysis, cartographic design, and information visualization to facilitate interactive, user-centered exploration of complex geospatial patterns (Liu et al., 2014). Web-based geovisualization tools have further democratized map-making, shifting control from institutional gatekeepers to communities and individuals, while enabling real-time interaction, multimedia storytelling, and participatory spatial analysis (Sui, 2004; Monmonier, 2015; Cartwright, 1999; Giordano, 2005). To meet increasingly sophisticated analytical needs, modern platforms now incorporate multivariate mapping, interactive dashboards, and coordinated-view systems that allow users to filter, link, and analyze spatial data across multiple visual representations (Giordano, 2005; Sarigai et al., 2025). These innovations, many of which are integrated into our *dciWebMapper2* framework presented in this paper, enhance analytical flexibility, encourage intuitive user engagement, and reveal patterns often obscured in static visualizations (MacEachren et al., 2014; Goodchild & Haining, 2004), setting the stage for more systematic, user-centered approaches to interactive map design.

Despite the advances in cartography, GIS, geovisualization, and web mapping, few systematic frameworks exist to guide the development of purpose-driven, user-friendly interactive web maps, or to support the professional use and creation of such maps in a consistent, replicable manner (Rinner, 2003; Sarigai et al., 2025). In response, *dciWebMapper* and the expanded version *dciWebMapper2,* offer structured, open-source approaches that unite the principles of cartography, GIS, and geovisualization with human-centered interaction design. The *dciWebMapper* and *dciWebMapper2* provide modular, extensible infrastructures for building web-based spatial applications that are not only technically rigorous but also accessible, inclusive, and adaptable across domains, from environmental justice to public health and transportation. Together, *dciWebMapper* and *dciWebMapper2* exemplify how interactive mapping can serve both expert and general users, bridging technical depth with interpretive clarity to support informed action and spatial understanding.

### 2.2 Enhancing user interaction: coordinated-view visualization and interface redundancy
Coordinated-view visualization has emerged as a foundational technique in both geovisualization and information visualization, enabling users to examine relationships across spatial and non-spatial dimensions by linking interactions across multiple views. These views (e.g., maps, charts, and data tables) are interconnected so that selections or filters applied in one automatically update the others, facilitating dynamic, multi-perspective exploration. Tools that implement coordinated views support exploratory workflows by allowing users to trace patterns, compare attributes, and uncover correlations that might otherwise remain obscured (MacEachren, 1995; Robinson, 2017; Kraak & Ormeling, 2020). Within geospatial data science, such techniques help bridge datasets across varying spatial scales and thematic domains, enriching analytical depth and enhancing user engagement (Dykes et al., 2005; Scherr, 2008; Robinson, 2017; Kraak & Ormeling, 2020). A more detailed overview is provided in Section 3.2 of the original *dciWebMapper* framework (Sarigai et al., 2025), where coordinated-view interaction is positioned as a cornerstone of the framework's design philosophy. The expanded framework builds on this foundation by implementing enhanced linking and filtering mechanisms across diverse map types and statistical charts, supporting more intuitive spatial reasoning and cross-domain insight.

Complementing these coordinated interactions, the expanded framework, *dciWebMapper2,* also emphasizes *interface redundanc*y, an important strategy for improving usability, accessibility, and inclusivity in open-source geovisualization tools. Interface redundancy refers to the practice of offering multiple, complementary pathways for accessing and interpreting the same information across different interface elements (Tindall-Ford et al., 1997). Interface redundancy is closely related to the concept of information redundancy, which reinforces understanding by repeating or re-encoding content in different formats (e.g., textual annotations paired with visuals) (Trypke et al., 2023). In the context of user-centered design (UCD), interface redundancy acknowledges that users have varied cognitive and perceptual preferences (Koh et al., 2025). Some users may process information more effectively through textual descriptions, while others may rely on visual cues such as maps or charts, and even within those groups who prefer visual cues, preferences for chart types or map styles may differ (Bertin, 2010). By offering strategically redundant elements (e.g., a status bar for current selection, synchronized dynamic highlights across multiple views, and explanatory tooltips), the expanded framework, *dciWebMapper2,* supports a broader range of user needs and promotes more equitable engagement. This commitment to redundancy is particularly important in public-facing and educational geovisualization tools, where accessibility and clarity are essential for fostering inclusive participation and meaningful interpretation of complex geospatial information.

**2.3 Open source and open science**

Open-source development and open science practices are foundational for promoting transparency, reproducibility, and equitable access in geospatial research (Brunsdon & Comber, 2021; Wang et al., 2025). The *dciWebMapper* framework (Sarigai et al., 2025) and its expanded version *dciWebMapper2* are intentionally released as open-source to support collaboration, innovation, and broader accessibility, including beyond institutional usage. Built with modular architecture and widely supported libraries (e.g., Leaflet.js, D3.js, DC.js, Crossfilter.js, Turf.js), the *dciWebMaper2* framework supports flexible adaptation for diverse analytical needs. While openness expands access, it does not guarantee usability or inclusiveness, both of which require deliberate design strategies. A key motivation behind the development of *dciWebMapper* and *dciWebMapper2* is to ground open-source, web-based mapping tools in the principles of cartographic design and information visualization, ensuring that visualizations are not only technically robust but also communicatively effective. As explored in a related study (partial authors of this paper, in preparation), UCD further complements this approach by aligning tool functionality with the needs of diverse users, including those with limited technical backgrounds (Emma, 2024; Slocum et al., 2001). UCD principles, utilized in *dciWebMapper2*, such as iterative feedback and interface redundancy, can extend the accessibility of open-source tools, enhancing their potential for public engagement, civic participation, and equitable decision-making.

### 3. *dciWebMapper2*: a significant extension to the initial *dciWebMapper* framework

The *dciWebMapper2* builds on the original framework (Sarigai et al., 2025), reusing its modular architecture and core design principles, coordinated-views, integration of spatial and non-spatial data, and a layout design that retains the original configuration while expanding it to include two enhanced map layout options. Rather than replacing the original *dciWebMapper*, it extends its functionality with additional map and chart types, including bivariate visualizations. Sections 3.1 to 3.8 detail new layout options, advanced mapping features, expanded charting capabilities, data structures, interactive enhancements, Leaflet-based user interface (UI) components, map reprojection support, and other key improvements. These components work together to enable cross-filtered coordinated views across maps, charts, and data tables, supporting interactive and exploratory geospatial data analysis. Reflections on how these extended functionalities align with the original design principles are presented in Section 4.5.

### 3.1. Map layouts

In *dciWebMapper2*, we expand upon the original single-map-with-charts layout from *dciWebMapper* by introducing two new layout options. The first new layout remains a single large, screen-width map (Figure 1), while the second supports multiple maps, each paired with corresponding charts (Figure 2), making them well-suited for visualizations with numerous layers and interactive functions, while supporting both readability and usability. These two layouts allow users to organize maps and visualizations in alignment with their datasets and analytical objectives. For detailed guidance on selecting appropriate map layouts, please refer to the "Maps/Charts Guide and Related Resources" page on any of the three web map applications(e.g., https://geoair-lab.github.io/iNMsocialJusticeMap/relatedResources.html#section4).

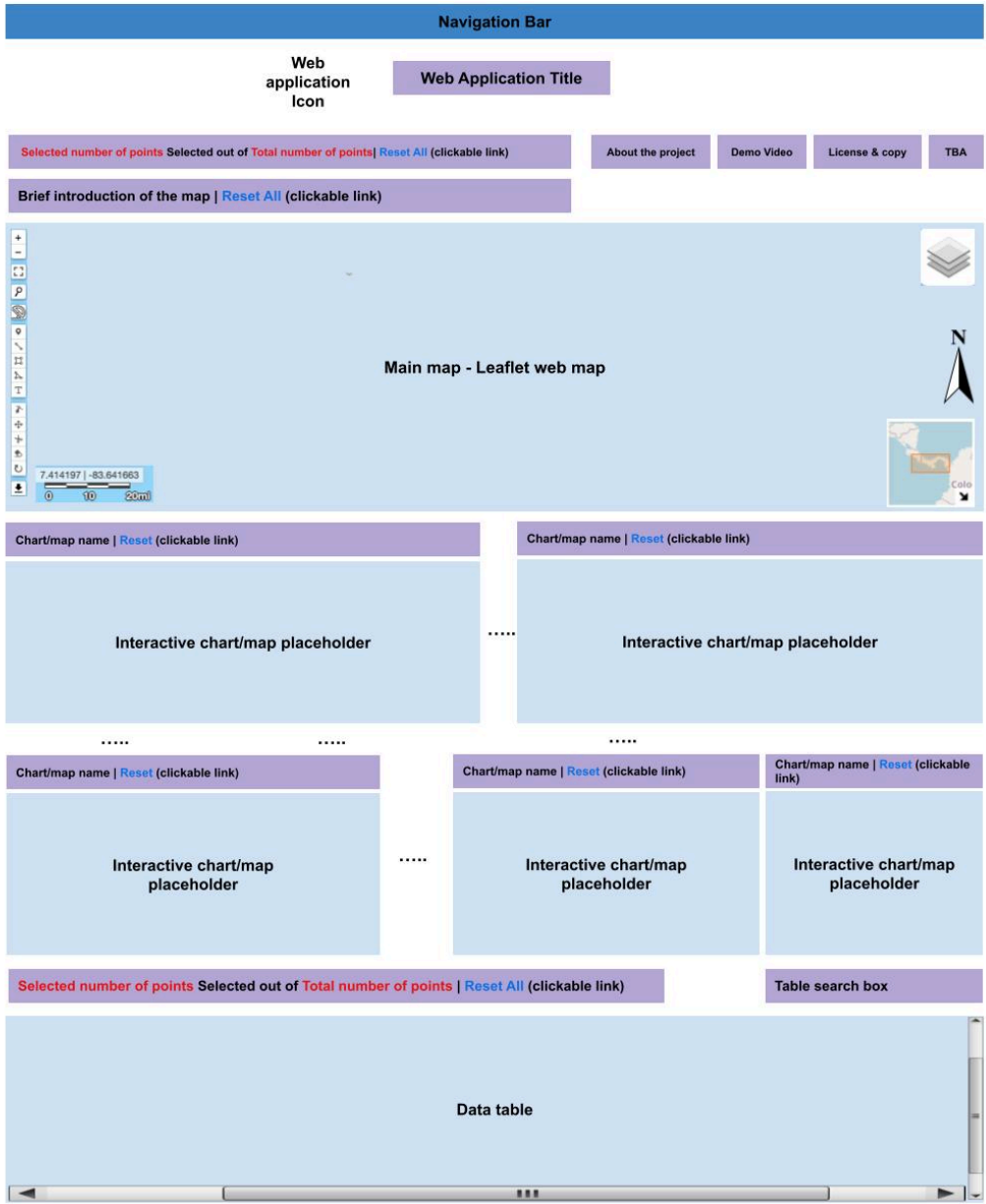

Figure 1. Single main map layout with a fixed navigation bar and an anchored status bar. This layout is designed for geospatial datasets with a horizontally elongated geographic extent (east-west-oriented geography), such as the iPathogenTrackingMap discussed in Section 4.1. The study area in the example, Panama, features a narrow, latitudinally stretched geography, making this layout ideal for effectively visualizing such regions. The single main map layout ensures that the horizontal stretch of the data area is fully utilized, enhancing spatial readability and supporting efficient visual spatial analysis. A fixed top navigation bar allows users to quickly and intuitively navigate between key interface sections: the main map row, several rows of maps/charts, and a concluding data table row.

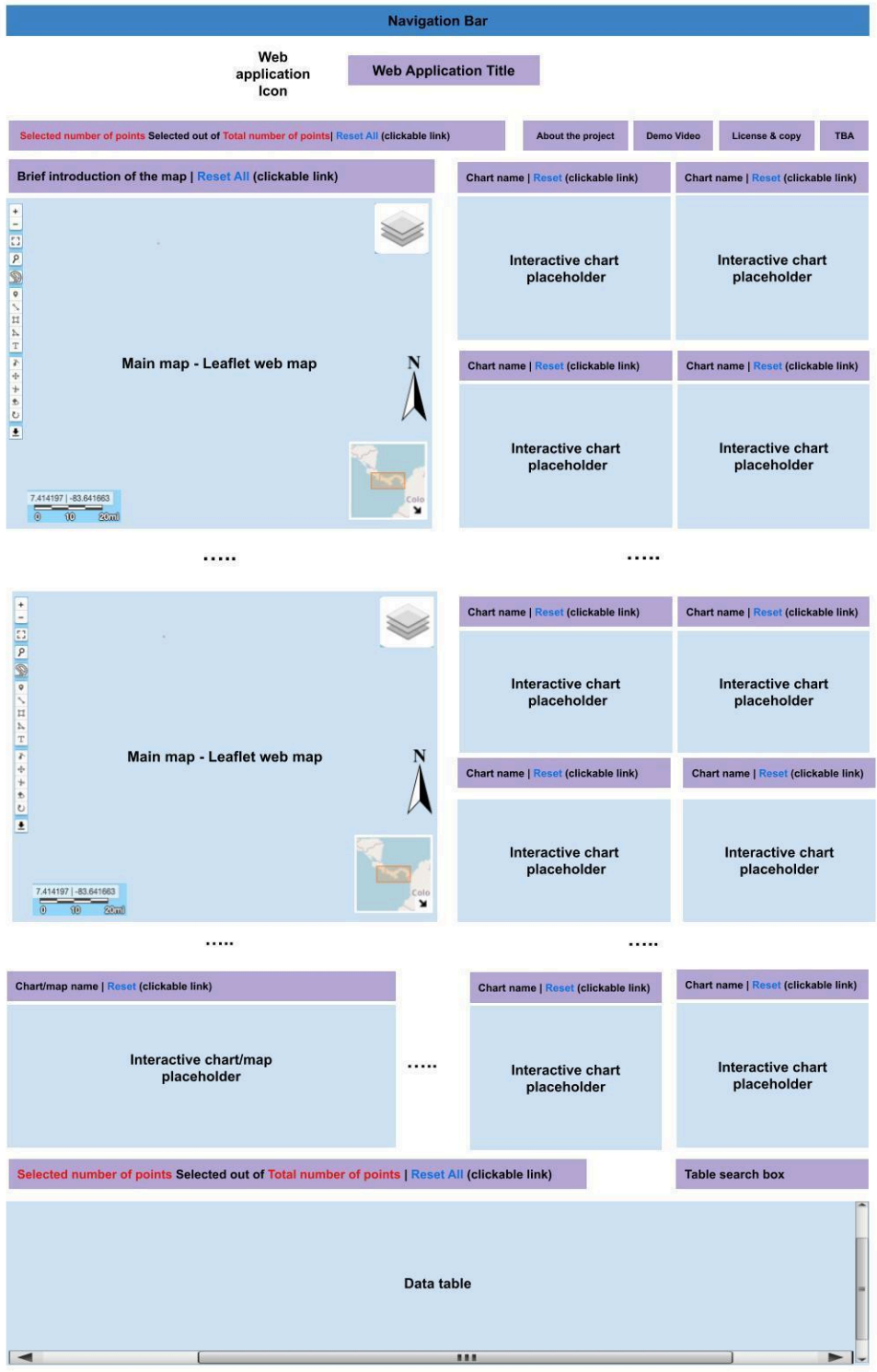

Figure 2. Multiple main-maps layout. This layout demonstrates the flexibility of *dciWebMapper2* in supporting multiple main maps, each accompanied by its own set of associated charts. Each map row represents a distinct geospatial visualization, while the corresponding charts display related data insights. Ellipses "..." are used between sections to indicate that additional map–chart

groupings can be dynamically added, depending on the number and complexity of datasets. This scalable design allows users to tailor the interface to diverse visualization needs, enabling seamless integration of geospatial data and interactive visual analytics.

### 3.2. Extended map types

Our *dciWebMapper2* framework extends the original by incorporating additional map types—proportional symbols, choropleths, small multiples, and heatmaps—beyond the original point-based marker maps. Dropdown-enabled thematic choropleth mapping allows dynamic variable comparison, supporting diverse geospatial analysis needs for effective exploration of complex spatial patterns.

Univariate, bivariate, and multivariate maps are foundational techniques in thematic cartography, each offering distinct approaches for visualizing and interpreting spatial patterns across one or more variables (Nelson, 2020; Slocum et al., 2022). ***Univariate maps*** visualize a single variable measured across discrete spatial units, with each location associated with one attribute value (Haining, 1981). Haining formalized univariate mapping within spatial statistics and GIS, establishing a theoretical foundation for analyzing spatial distributions in modern applications. Historically, examples include John Snow's 1854 cholera map and thematic works by von Humboldt and Minard, which illustrated variables such as mortality, elevation, and trade (Snow, 1854; Robinson, 1982; Chen et al., 2008). Univariate maps effectively communicate spatial patterns, especially to non-technical audiences. However, they may oversimplify complexity, obscure local variation through aggregation, and mislead if poorly designed (Haining, 1981). Frameworks like *dciWebMapper* and *dciWebMapper2* address these challenges by embedding thoughtful cartographic design and interaction principles that help users produce clear, interpretable visualizations. Bivariate and multivariate maps further enhance spatial insight by enabling richer comparisons across multiple variables. ***Bivariate and multivariate maps*** encode two or more variables within a single visualization, revealing relationships often missed in univariate representations (Nelson, 2020). Their core purpose is to compare the spatial distribution of multiple phenomena, efficiently conveying more information in a compact format (Slocum et al., 2022; Kraak et al., 2020; Nelson, 2020; Korycka-Skorupa & Gołębiowska, 2021). While they can highlight both anticipated and unexpected correlations, increased visual complexity may hinder interpretation, particularly for general audiences. High cognitive load, design challenges, and data compatibility issues can reduce their effectiveness. Thoughtful design and, when appropriate, complementary techniques like small multiples or interactive tools can improve clarity and usability.

While bivariate and multivariate maps offer simultaneous multivariable insight, *dciWebMapper2* focuses on several extended univariate map types to enhance interpretability, flexibility, and user accessibility, as detailed in Sections 3.2.1 to 3.2.4, including choropleths, proportional symbols, heatmaps, and small multiples.

3.2.1 Choropleth maps

*Choropleth maps* are fundamental tools in thematic cartography, used to visualize aggregated numerical data, such as rates or densities, across geographic units through shading or color (Andrienko et al., 2001). Originating with Dupin's 19th-century map of education in France, they are widely used to reveal spatial patterns and trends. Choropleths are **best suited for normalized numerical data, not categorical variables, as normalization (e.g., converting raw totals to rates) ensures comparability and prevents misleading representations** (Schiewe, 2019;

Monmonier, 2001; Slocum et al., 2022). Despite their utility, they may obscure local variation due to data aggregation, are sensitive to scale, and can mislead if poorly designed (Robinson, 1978; Jenks, 1963). Alternative techniques, such as proportional symbols (Section 3.2.2) and dot density maps, can address these limitations by offering more nuanced representations of spatial data (Slocum et al., 2022).

### 3.2.2 Proportional symbol maps

*Proportional symbol maps* represent data by scaling symbols such as circles, bars, or pictograms based on raw totals (Golebiowska et al., 2021; Slocum et al., 2022). Unlike choropleth maps, which use sequential color gradients to depict standardized rates across enumeration units (Brewer, 2016), proportional symbols highlight absolute magnitudes. This distinction matters, as high rates may mislead without sufficient base counts (Slocum et al., 2022). Proportional symbol maps also support bivariate mapping by encoding a second variable through color, enhancing multidimensional insight. While they can become *cluttered in dense areas, our framework integrates them with interactive charts, enabling click-based filtering and dynamic updates for exploratory analysis*.

### 3.2.3 Heatmap

Originally introduced by Kinney (1993), *heatmaps* are widely used to visualize spatial patterns in high-volume point data (Netek & Slezakova, 2018; Sun & Li, 2013; Zhang et al., 2020). Using color gradients to show intensity helps reveal clusters, trends, and outliers. GIS heatmaps include point density and weighted types, the former based on concentration, the latter incorporating attribute values (Netek & Slezakova, 2018). While intuitive, heatmaps require careful design; *inappropriate color ramps, blur radius, or lack of normalization can distort interpretation.* Unlike choropleth or proportional symbol maps, heatmaps convey relative intensity rather than exact values. Our *dciWebMapper2* integrates both global (static) and local (interactive) point density heatmaps to visualize crash incidents in Albuquerque (see Figure 14), enabling users to explore overall trends and dynamically respond to filtered data, supporting both broad and targeted multiscale analysis (Pokojski et al., 2021).

### 3.2.4 Small multiples

Small multiples, introduced by Tufte (1983, 1991), are compact, consistent visualizations, like "postage-stamp-sized illustrations", that vary by category, time, or variable. They facilitate pattern recognition and comparison by aligning multiple views of the same dataset (Meulemans et al., 2016). This technique supports multivariate analysis by showing variable differences within a shared spatial or conceptual frame and can be applied across plots like scatterplots, bar charts, and choropleth maps (Shkolnik, 2017). In our *dciWebMapper2*, we implement interactive small multiples using choropleth maps. Users can click to explore data; selections in one map dynamically and simultaneously update related charts and maps, enabling coordinated multiscale spatial analysis and enhancing exploratory insight.

### 3.3 Extended chart types

In geovisualization, charts complement maps by supporting interactive knowledge-building (Dykes et al., 2005). Charts can be static or dynamic (Chen et al., 2009); *dciWebMapper2* adopts dynamic charts to enhance user engagement and streamline insight retrieval (Knaflic, 2015). The original *dciWebMapper* includes bar charts, line charts, pie/donut charts, row charts (including scrollable X/Y axes), sunburst charts, word clouds, and data tables, offering diverse pathways for data exploration (Sarigai et al., 2025). *dciWebMapper2* expands this chart set, increasing flexibility to match visual formats with analytical needs. While guidance on original chart types is provided in Section 3.3 (Sarigai et al., 2025), this paper introduces extended chart types, along with recommendations for effective use.

Univariate and bivariate charts are foundational tools in data visualization, enabling users to explore the distribution of individual variables and the relationships between pairs of variables, respectively. *Univariate charts* show the distribution of a single variable, revealing a central tendency, variability, and overall data shape (Cleveland, 1993). Bar and pie charts are used for categorical data, while histograms and boxplots suit quantitative variables. These foundational tools support descriptive analysis by helping identify patterns, outliers, and data quality issues. *Bivariate charts* visualize relationships between two variables, helping reveal patterns, trends, or associations (Friendly, 2002; Wilkinson, 2011). Chart selection depends on the variable types (categorical or quantitative) and the analytical objective. These visualizations are central to exploratory analysis and hypothesis testing. Sections 3.3.1 to 3.3.5 introduce extended chart types implemented in our *dciWebMapper2* framework, with each type labeled as univariate and/or bivariate in parentheses.

#### 3.3.1 Histogram (Univariate)

*Histograms* visualize the distribution of quantitative data by grouping values into intervals (bins) and displaying their frequency as bar heights. The choice of bin size is crucial: small bins may introduce noise, while large ones can obscure patterns. Widely used in measurement and simulation analysis, histograms help identify trends and distributions (Scott, 1979; Freedman & Diaconis, 1981). In *dciWebMapper2*, we implement interactive histograms that allow users to adjust bin size via a dropdown and filter data through brushing. These features dynamically update linked visualizations, making histograms a powerful tool for EDA.

#### 3.3.2 Range Slider (Univariate)

A **Range Slider** is an interactive component that lets users select single values or continuous ranges along a scale, facilitating dynamic filtering and rapid numeric parameter adjustment in EDA. Our *dciWebMapper2* implements this feature using the open-source library *ion.RangeSlider.js*[1].

#### 3.3.3 Stacked bar charts (Bivariate)

*Stacked bar charts*, an extension of standard bar charts in *dciWebMapper*, display numeric values across two categorical variables by dividing each primary bar into sub-bars representing secondary categories. This design helps users compare totals and understand category composition (Knaflic, 2015). Especially useful for analyzing proportional distributions, stacked bars reveal how totals are broken down across subgroups. *dciWebMapper2* allows users to filter data interactively by selecting sub-bars, with real-time updates reflected across all linked maps and charts.

3.3.4 Scatterplots (Bivariate)

*Scatterplots* effectively visualize relationships between two quantitative variables, helping reveal patterns, trends, or correlations (Moore et al., 2013). By plotting data points along two axes, they are especially useful in exploratory and correlational analysis. In our framework, users can brush (select) specific plot regions to filter data in real-time, with updates reflected across all linked charts and maps. This interactive feature aids in spotting relationships and anomalies within large datasets. The SVG (Scalable Vector Graphics)-based design ensures smooth, responsive performance for scalable and interactive visual analysis.

**Scatterplots with regression lines** enhance the visualization of relationships between two variables by combining raw data points with a trendline (best-fit or least-squares) that clarifies patterns and supports prediction. While trends may be linear or nonlinear, regression lines simplify interpretation, especially for exploratory analysis (Anscombe, 1973; Cleveland & Devlin,1988). In *dciWebMapper2*, users can dynamically select x- and y-axes via dropdowns, allowing flexible investigation of variable correlations. This interactive design aligns with data exploration best practices, helping uncover meaningful insights.

3.3.5 Boxplots (Univariate, grouped bivariate)

*Boxplots* are compact visual tools for summarizing univariate data distributions, showing the median, interquartile range, minimum, maximum, and outliers (Frigge et al., 1989). They are ideal for assessing variability, symmetry, and identifying outliers, and are especially useful for comparing distributions across groups. While traditionally univariate, boxplots can also support bivariate analysis when grouped by a categorical variable. In *dciWebMapper2*, interactive boxplots allow users to select ranges or outliers, dynamically updating all connected maps and charts in real-time. This interactivity supports coordinated, exploratory analysis of complex datasets.

### 3.4. Extended data table functionality

The data table remains a core component of *dciWebMapper2*, offering users a versatile interface for exploring complex datasets. As Knaflic (2015) notes, tables effectively engage the verbal system and are often the clearest way to convey information. Like in the original *dciWebMapper*, the table supports overview, filtering, sorting, search, and details-on-demand, while allowing many columns to present rich semantic context. The major enhancement in *dciWebMapper2* is *bidirectional interactivity*: users can now filter maps and charts by clicking rows in the table, reversing the original one-way updates. This seamless coordination improves exploratory workflows across tabular and spatial views. For point-based data (e.g., iPathogenTrackingMap, Section 4.2), the clickable filtering functionality in the data table enables precise filtering and geographic identification of selected samples. For aggregated data (e.g., choropleth maps in iNMsocialJusticeMap in Section 4.3), the data table's search function may be more appropriate than its clickable filtering, as it supports focused lookup without disrupting broader visual context provided by charts and maps.

## 3.5. Data structures and types for extended maps and charts

Effective visualization requires aligning visual components with appropriate data structures and types. In *dciWebMapper2*, extended maps, charts, and data tables rely on structured inputs that guide rendering and interactivity. This section details the data formats supporting these components, emphasizing the importance of selecting map and chart types suited to the dataset's structure. Tables 1 and 2 summarize data requirements, legend use, interaction types, and operators for each component. 'NA' indicates non-applicable visual options.

Table 1. The data type and functions of the map and map-related components.

| Map/map-related component | Data Type | Legend | Events/Interaction | Interaction Operator |
|---|---|---|---|---|
| Marker map | Location data (point) | Marker map legend | Click marker, zoom in/out, spiderfy clusters | Retrieve, zoom, search, filter, pan |
| Choropleth map | Area-based data (e.g., tract) | Color gradient legend | Hover, click area, zoom in/out, layer switch | Retrieve, zoom, search, filter, resymbolize, pan |
| Proportional symbol map | Quantitative point data | Size-based legend | Hover, click symbol | Retrieve, search, zoom, filter, pan |
| Small multiple maps | Multi-variable, area-based | NA | Compare, hover, click area | Retrieve, zoom, filter, reproject, arrange, sequence, reexpress, pan |
| heatmap | High-density point data | NA | Mouse move, zoom | Retrieve, zoom, resymbolize, zoom, pan |
| Basemap | Tile layer | NA | Turn on/off | Retrieve, zoom, overlay, reexpress |
| Pop-up | Text, image, and audio data | NA | Scrollable, content can be clickable | Retrieve |

Note: *NA* refers to not applicable. The interaction operator *"resymbolize"* refers to modifying a map's design parameters, such as color, size, or classification, without changing its fundamental type. In contrast, *"reexpress"* involves switching the form of representation (e.g., from a choropleth to a proportional symbol map) while preserving the underlying data. See Roth (2017) for a comprehensive overview of interaction operators and their definitions.

Table 2. The data type and functions of extended charts and data table in *dciWebMapper2*.

| Chart/table | Data Type | # of Variables | Legend-related Events/ Interaction | Events/ Interaction | Interaction Operator |
|---|---|---|---|---|---|
| Stacked bar chart | Categorical data | Bivariate | Mouseover | Clickable, filterable, mouseover, mouseout | Filter, retrieve |
| Scatterplot | Continuous data | Bivariate | Mouseover | Brushing, clickable, filterable | Filter, retrieve |
| Boxplot | Continuous data | Univariate (per category), bivariate | Mouseover | Clickable, array filter/dimension tag, filterable, mouseover, mouseout | Filter, retrieve |
| Scatterplots with regression lines | Continuous data | Bivariate | NA | x-axis scroll, y-axis scroll, brushing, array filter/dimension tag, filterable, mouseover, mouseout | Filter, retrieve |
| Histogram | Continuous | Univariate | NA | Clickable, x-axis scroll, y-axis scroll, array filter/dimension tag, filterable, mouseover, mouseout | Filter, retrieve |
| Range slider | Continuous data | Univariate | NA | Clickable, filterable | Filter, retrieve |
| Select menu | Categorical data | multivariate | NA | Clickable, filterable | Filter, retrieve |
| Data table | Continuous data, categorical data | multivariate | NA | Searchable, paginated, horizontally scrollable, clickable links in table, ordering, clickable row to update other charts/map | Search, sequence, retrieve, sort |

Note: *NA* refers to "not applicable." Brushing is an interactive technique that allows users to define a one- or two-dimensional selection range, typically by clicking and dragging the mouse over a chart or map, to highlight, filter, or link data points across coordinated views. This feature supports exploratory analysis by enabling users to focus on specific subsets of data while maintaining visual context. The "array

filter/dimension tag" in dc.js enables filtering based on columns that contain multiple comma-separated values, essentially allowing a single data field to represent multiple categorical attributes. For example, in the iPathogenTrackingMap, the "part" column may include entries such as "blood, liver, lung" or "blood, liver, whole organism". When this column is used as a dimension in a chart, dc.js can parse and treat each item in the list (e.g., "blood", "liver", "lung") as an individual filterable value, rather than as a single compound string. This means users can interactively filter data based on any combination of individual terms within these lists, greatly enhancing the flexibility and granularity of exploratory analysis. This advanced capability supports more realistic, multifaceted classification and tagging of records, which is particularly useful in domains where entities (e.g., samples, regions, species) are associated with multiple categories simultaneously.

**3.6 Extended mapping, interaction, and UI functionality integrated through Leaflet plugins**
Leaflet is a powerful open-source JavaScript (JS) library for interactive maps, made more versatile through plugins that extend functionality and support advanced, customized geospatial visualizations.
For details on what key Leaflet plugins are leveraged and how they enhance the overall capabilities of the *dciWebMapper2* framework, please refer to the Table on the "Charts/Maps Guide and Related Resources" page on any of the three web map applications (e.g., https://geoair-lab.github.io/iNMsocialJusticeMap/relatedResources.html#section5)

**3.7. Integrating reprojection functionality**
Map projections are essential in cartography, each balancing distortions of shape, area, distance, or direction to suit different purposes. The Robinson projection offers a visually balanced compromise for world maps (Lapaine & Frančula, 2022), while the Mercator projection, used by Google Maps, OpenStreetMap, and Mapbox, is conformal and navigation-friendly but distorts area (Battersby et al., 2014). Web Mercator simplifies Mercator by mapping the Earth to a sphere, enabling fast tile-based rendering and continuous zooming. Our *dciWebMapper2* highlights the importance of selecting map projections based on the specific purpose of each map. For area-accurate choropleth maps, equal-area projections are crucial. To support precision, we incorporate a projection selection feature, enabling users to define appropriate projections for their visualizations. We integrate three projection types: (1) Albers' equal-area conic, for preserving area in regional thematic maps; (2) Cylindrical (e.g., equirectangular, spherical Mercator), for simplicity and web compatibility; (3) Azimuthal stereographic, for conformal properties in polar or hemispheric views. These options enhance spatial accuracy and flexibility across diverse mapping needs.

**3.8. Significant enhancements**
The *dciWebMapper2* introduces several significant enhancements that boost spatial interactivity, enrich attribute exploration, and integrate lightweight spatial analysis, expanding both analytical depth and user experience. *(1) Leaflet MarkerCluster spiderfy.* Overlapping point markers, especially common in datasets with repeated sampling at the same location (e.g., pathogen hosts; see Section 4.2), can clutter maps and limit usability. To address this challenge, *dciWebMapper2* integrates the Leaflet.MarkerCluster plugin (Section 3.6), which dynamically groups nearby markers into clusters based on zoom level. As users zoom in, clusters separate, and overlapping markers are "spiderfied" for clear, individual selection. This elegant spiderfy functionality enhances clarity, interactivity, and navigation, facilitating smooth exploration even in dense datasets. Figure 3a illustrates the implementation. *(2) Expanding attribute exploration through* *dropdown-driven choropleth mapping.* The *dciWebMapper2* introduces dropdown-driven choropleth mapping to enable dynamic, multi-attribute exploration across spatial themes. In the iNMsocialJusticeMap case study (Section 4.4, Figures 6–8), three choropleth maps, focused on climate and economic justice, Social Vulnerability Index (SVI), and food access, feature dropdown menus for switching between variables within each domain. This design supports in-map attribute comparison and cross-map thematic analysis, enhancing spatial reasoning and revealing how dimensions of vulnerability and equity intersect geographically. By allowing real-time updates of mapped attributes, this interface advances accessible, multilayered geovisualization for exploratory analysis. *(3) Use spatial analysis to filter data.* In *dciWebMapper2*, we introduced a spatial filtering function that allows users to interact directly with the map, drawing a polygon, radius, or bounding box, to filter data spatially. The selected area dynamically filters associated charts and tables, facilitating focused exploration of localized patterns and spatial relationships. This capability enhances geospatial analysis by supporting location-specific inquiries and real-time updates across visual components. To implement this, we build on core JS libraries used in the original *dciWebMapper* framework: D3.js[2], Crossfilter.js[3], DC.js[4], and Leaflet.js[5] alongside supporting tools like DataTables[6], jQuery[7], and Bootstrap[8]. The spatial filtering functionality relies on two key libraries: (a) Turf.js[9], for advanced spatial operations (e.g., checking whether points fall within drawn areas); and (b) Leaflet-Geoman[10], for user-defined geometry editing. Together, these enable an interactive spatial workflow where user-defined areas drive coordinated filtering across maps and charts. This feature is demonstrated in the iABQtrafficCrashMap case study (Figure 13, Section 4.4).

## 4. Case studies applying the *dciWebMapper2* framework and reflections on establishing design principles

To demonstrate the versatility of *dciWebMapper2* in developing interactive web map apps for high-dimensional geospatial big data analysis and visualization, we present three case studies: iPathogenTrackingMap (Section 4.2), iNMsocialJusticeMap (Section 4.3), and iABQtrafficCrashMap (Section 4.4). This section begins with an overview of the implementation, highlighting the core libraries and technologies used (Section 4.1), followed by in-depth descriptions of each use case (Sections 4.2–4.4), and concludes with reflections on the foundational design principles established in the original *dciWebMapper* framework (Section 4.5). Collectively, these case studies highlight the modularity and transferability of the original architecture established in *dciWebMapper* framework (Sarigai et al., 2025), showing how core principles from cartography and information visualization are effectively and meaningfully applied in diverse case studies of *dciWebMapper2*.

### 4.1. Implementation overview

The *dciWebMapper2* framework builds on the original *dciWebMapper* (Sarigai et al., 2025), which leveraged JS and core libraries, D3.js[2], Crossfilter.js[3], DC.js[4], and Leaflet.js[5], to support high-dimensional, interactive, and coordinated-view enabled geovisualization. To expand its capabilities, *dciWebMapper2* integrates additional open-source tools that enable more dynamic user interaction and advanced spatial analysis. Leaflet-Geoman[10] allows for in-map drawing and editing (e.g., selecting census tracts), while Turf.js[9] performs client-side spatial operations such as geometric intersection and attribute summarization. A suite of Leaflet plugins (see Section 3.6) further enhances usability with features like minimaps, measurement tools, and extended layer controls. The framework retains a modular, extensible JS architecture and is deployed as open-access web apps hosted on GitHub. See the DATA AND CODE AVAILABILITY STATEMENT for links to source code, brief demo videos, and datasets.

**4.2. iPathogenTrackingMap web map app**

4.2.1. Introduction, motivation, and data collection
Predicting pandemics requires integrating diverse data on pathogens, hosts, and human-environment interactions (Scarpino & Petri, 2019; Colella et al. 2023). Georeferenced samples offer unique insights into the spatial and temporal dynamics of pathogen communities in natural settings (Azat et al., 2024; Casanovas-Massana et al., 2018; Walerius et al., 2023). Mapping pathogen-host relationships supports early detection of high-risk areas, clarifies environmental and social drivers of outbreaks, and enhances data-driven decision-making for public health.

Our iPathogenTrackingMap web map app visualizes hantavirus screening data from small mammals collected in Panama (1990–2020) by the Instituto Conmemorativo Gorgas de Estudios de la Salud (Armién et al. 2023; Gonzalez et al. 2023) and curated and made available by the Museum of Southwestern Biology[11], including GPS locations, collection dates, and Choclo virus (CHOV) results. Host bio-information is linked via Arctos (Cicero et al., 2024), and socioeconomic variables, from Panama's Census Bureau and STRI accessed from the GIS data portal[12], cover healthcare access, sanitation, population demographics, and infrastructure. Using coordinated-view geovisualization, our iPathogenTrackingMap combines maps, charts, and tables for dynamic exploration of spatial and temporal patterns. By integrating pathogen and socioeconomic data, the app enables users to detect trends, examine correlates of disease transmission, and support proactive surveillance and intervention strategies.

4.2.2. Web app design and implementation
The iPathogenTrackingMap web app, accessible at https://geoair-lab.github.io/iPathogenTrackingMap/index.html, visualizes georeferenced Hantavirus host data from small mammals collected in Panama (1990–2020) and archived at the Museum of Southwestern Biology. The dataset includes GPS coordinates, scientific names, collection metadata (location, date), biological traits (e.g., sex, weight), and Orthohantavirus screening results. Socioeconomic layers, covering healthcare, sanitation, and population demographics, are integrated from the Panama Census Bureau and the Smithsonian Tropical Research Institute. Following *dciWebMapper* (Sarigai et al., 2025) and *dciWebMapper2* design principles (see Sections 3.2–3.4 for *dciWebMapper2* extended map and charts, and data table functionality). The cartographic and information visualization design principles established in *dciWebMapper* are in Section 4.1 to Section 4.2 of Sarigai et al. 2025); and a brief outline and description of those design principles is also available at https://geoair-lab.github.io/iPathogenTrackingMap/Carto&InfoVisDesignPrinciples.html. We implemented multiple coordinated-view visualizations for dynamic exploration of spatial and temporal patterns.

**Map visualizations (Figure 3).** 3(a) *Marker map*: Shows spatial distribution with symbols colored by screening results. Clustered markers improve readability and interactivity (Leaflet plugin). 3(b) *Choropleth map*: Displays positive sample rates by state; selections update all linked views. 3(c) *Proportional symbol map*: Visualizes cumulative sample counts per state.

These complementary map types promote interface redundancy and support varied analytical tasks.

**Charts and interactive features.** (1) *Donut charts* for screening outcomes and host sex (Figure 4a–b). (2) *Row charts* for body parts examined, preferred over pie charts for clarity with many categories. (3) *Date slider* to filter by collection time (Figure 5a). (4) *Boxplot* for examining host gender against numeric variables (Figure 5b).  (5) *Scatterplot* with dropdown-controlled axes and brushing to highlight patterns (Figure 5c–d). (6) *Stacked bar chart* for species and screening results (Figure 5e). (7) *Series line chart* tracking weight of positive/negative hosts over time (Figure 5f). (8) *Select menu* for filtering by scientific name (Figure 5g).

All views are fully interactive and coordinated, with real-time updates on user interaction. Key functionalities include leaflet cluster markers, informative pop-ups, toggleable layers, and a mini-map for navigation. A short demo video is embedded in the app's top-right corner. For additional data structure and variable descriptions, visit the detailed documentation page https://geoair-lab.github.io/iPathogenTrackingMap/dataStructureExplaination.html.

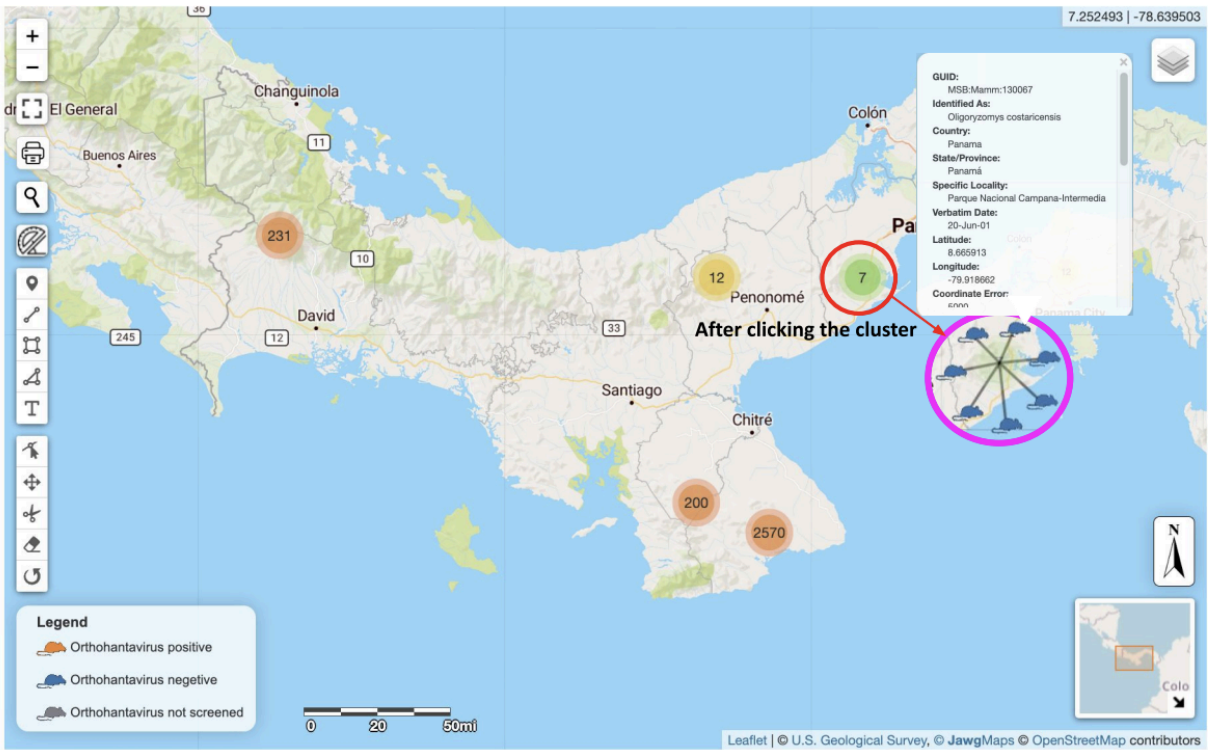
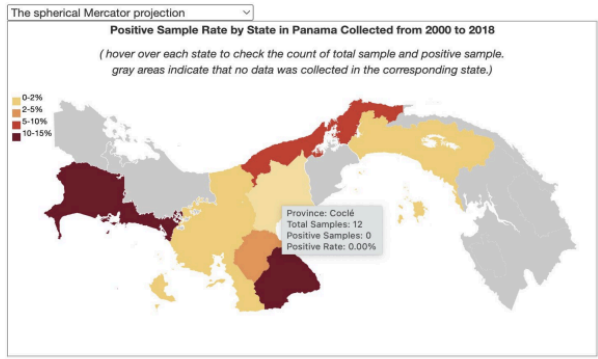
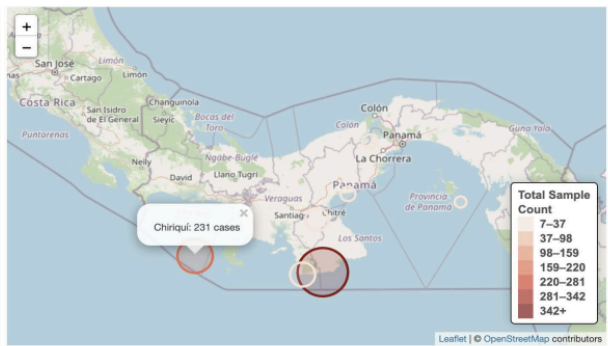

Figure 3. Geospatial visualizations of host collection in iPathogenTrackingMap. (a) Sample distribution map with Leaflet Marker Clusters showing the spiderfy effect in *dciWebMapper2*. Overlapping markers at the same location (e.g., multiple pathogens on a single host) are clustered (green cluster marker indicated with the red circle) and expand radially on click (pink circle), enabling inspection of individual sample records via interactive popups. (b) Choropleth map displaying positive sample rate by state. (c) Proportional symbol map illustrating cumulative sample counts per state. Note: A brief web map app demo video is accessible on the iPathogenTrackingMap page and applies to the interactive features illustrated in Figures 3–5 (see DATA AND CODE AVAILABILITY STATEMENT).

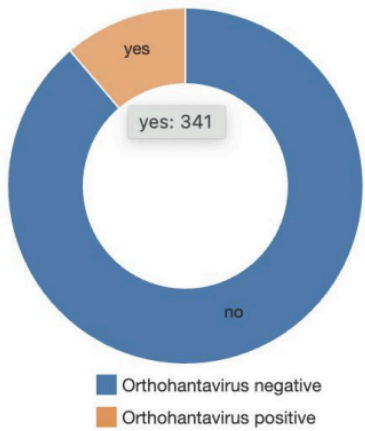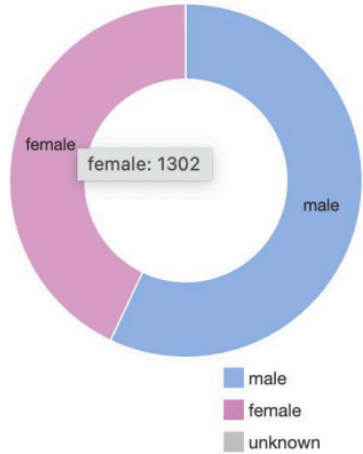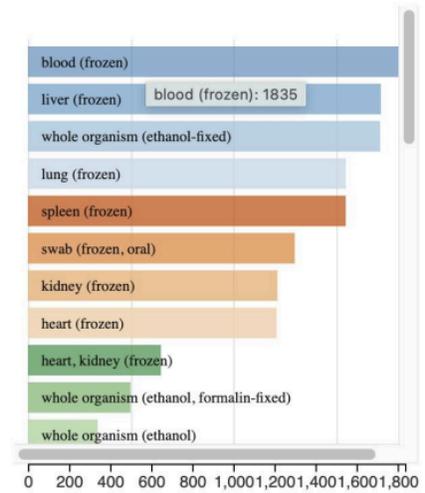

Figure 4. Donut and row charts for summarizing host attributes in iPathogenTrackingMap. (a) Orthohantavirus screening results (donut chart), (b) host sex distribution (donut chart), and (c) sampled body parts (row chart). Donut charts include scrollable legends; row charts support scrollable axes for improved readability.

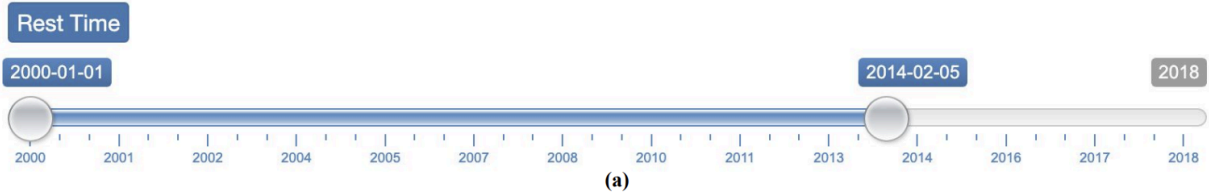

(a)

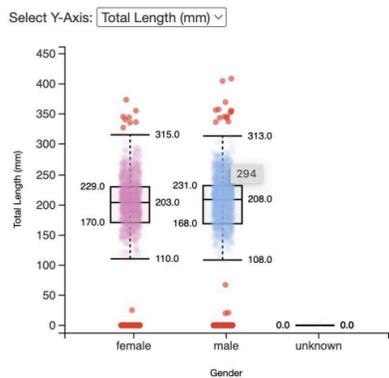

(b)

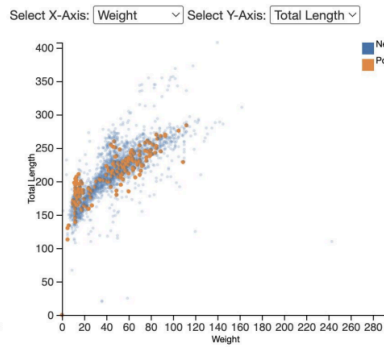

(c)

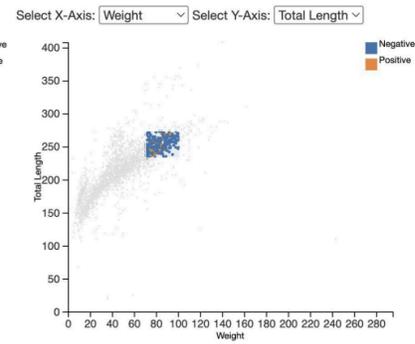

(d)

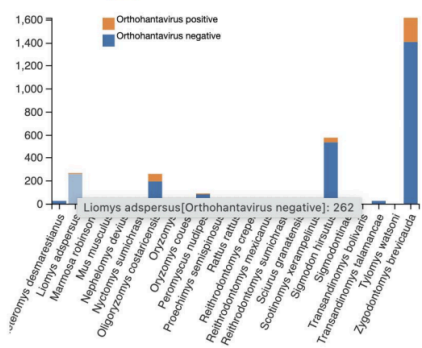

(e)

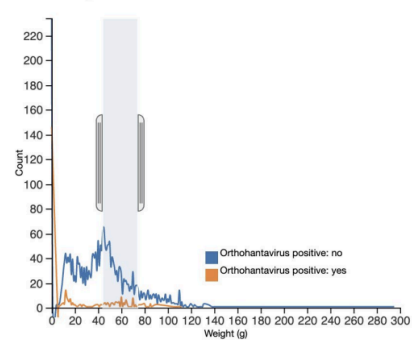

(f)

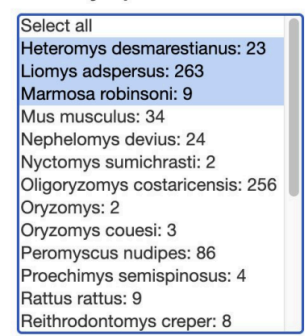

(g)

Figure 5. Interactive visualizations in iPathogenTrackingMap. (a) *Date range slider* filters samples by year, month, and date. (b) *Boxplots* shows host gender vs. a user-selected variable, with clickable points for cross-filtering. (c) *Scatterplot* visualizes relationships between two variables; point colors indicate Orthohantavirus results. (d) Polygon-brushing highlights selected scatterplot points in (c). (e) *Stacked bar chart* categorizes scientific names by screening results. (f) *Series line chart* displays host weight trends by result. (g) *Select menu* enables filtering by scientific name (top three are selected).

### 4.3. iNMsocialJusticeMap web map app

4.3.1. Introduction, motivation, and data collection

Addressing systemic inequalities requires integrated justice frameworks, social, economic, environmental, and food justice, and tools like the Social Vulnerability Index (SVI). *Social justice* promotes equitable access to resources and human dignity across race, gender, and income (Rawls, 1971; Bullard, 2005; Cadieux & Slocum, 2015). *Environmental justice* ensures fair treatment in decisions affecting health and ecosystems, especially for communities disproportionately exposed to harm (EPA, 2024). *Food justice* advocates for equitable access to nutritious, culturally appropriate food by addressing structural barriers like food deserts and affordability (Bennion, 2022; Murray et al., 2023). These social, environmental and food justice domains are interconnected: tackling food insecurity requires addressing socioeconomic disparities, for example, environmental justice demands attention to who bears environmental burdens (Cutter et al., 2003; Flanagan et al., 2011). Tools like the SVI help identify at-risk populations, enabling more targeted, equitable policy interventions.

Despite growing reliance on these frameworks, there's a pressing need for an integrated platform that unifies food, environmental, and economic justice indicators. This tool would reveal overlapping vulnerabilities, support holistic decision-making, and promote resilience in underserved communities. Our iNMsocialJusticeMap case study uses three key datasets at the census tract level in New Mexico: (1) SVI (CDC/ATSDR) for social vulnerability (Flanagan et al., 2011), (2) Climate and Economic Justice Screening Tool (CEJST) for environmental/economic indicators (CEQ, 2024), and (3) USDA Food Access Research Atlas for food justice metrics (USDA, 2025). All data were processed and integrated using Python, with census tract IDs as the common key, facilitating a comprehensive spatial analysis of justice-related challenges across the state.

4.3.2. Web app design and implementation
The iNMsocialJusticeMap web app (https://geoair-lab.github.io/iNMsocialJusticeMap/index.html) visualizes social, environmental, and food justice indicators at the census tract level in New Mexico, integrating datasets from CDC/ATSDR SVI (Flanagan et al., 2011), CEJST (CEQ, 2024), and the USDA Food Access Research Atlas (USDA, 2025). Using Python for preprocessing and *dciWebMapper2* for design guidance, the app supports multi-dimensional analysis of systemic inequities through intuitive, interactive visualizations.

**Map visualizations through three interactive choropleth maps (Figures 6–8) and *small multiple maps* (Figure 12).** *The three dropdown empowered interactive choropleth maps (Figures 6–8)* enable users to explore indicators across justice domains, with dropdown menus for customizing views. These maps cover: Climate & Economic burdens (e.g., pollution, housing, workforce), SVI indicators (e.g., socioeconomic status, minority status), and low-income and low-access food areas. *The small multiple maps* illustrated in Figure 12 demonstrate how projection choices affect spatial interpretation. Users can compare demographic and socioeconomic patterns across the default and Albers projections, with dynamic map-linked interactions.

**Charts and interactive features.** A suite of charts (Figure 9) enhances comparative insights: (1) *Donut charts* show distribution by county, burden categories, disadvantaged tracts, and food access (1, 10, 20 miles; ½ mile without vehicle). (2) *Bar charts* display low-income tracts and Supplemental Nutrition Assistance Program (SNAP) usage. (3) Further interactivity is achieved through *selection menus* and *histograms* (Figure 10), enabling tract-level analysis and dynamic filtering across poverty, socioeconomic status, household characteristics, minority status, and housing/transportation. (4) ***Three interactive scatterplots with regression lines*** (Figure 11) support comparative analysis by allowing users to select variables from justice datasets via dropdown menus, generating scatterplots with trend lines to explore bivariate relationships. Each interactive scatterplot features polygon-based brushing, enabling users to highlight specific subsets, such as outliers or points near the trend line, for focused analysis. This interactive design reveals spatial and statistical patterns not accessible in static charts, helping users intuitively explore relationships and uncover insights into associated environmental and social vulnerability attributes.

The app includes a demo video (accessible from the top-right corner of iNMsocialJusticeMap) and a detailed data structure guide (https://geoair-lab.github.io/iNMsocialJusticeMap/dataStructureExplaination.html), supporting transparency and user onboarding. This design reflects a commitment to clarity, usability, and the integration of justice-focused spatial analysis.

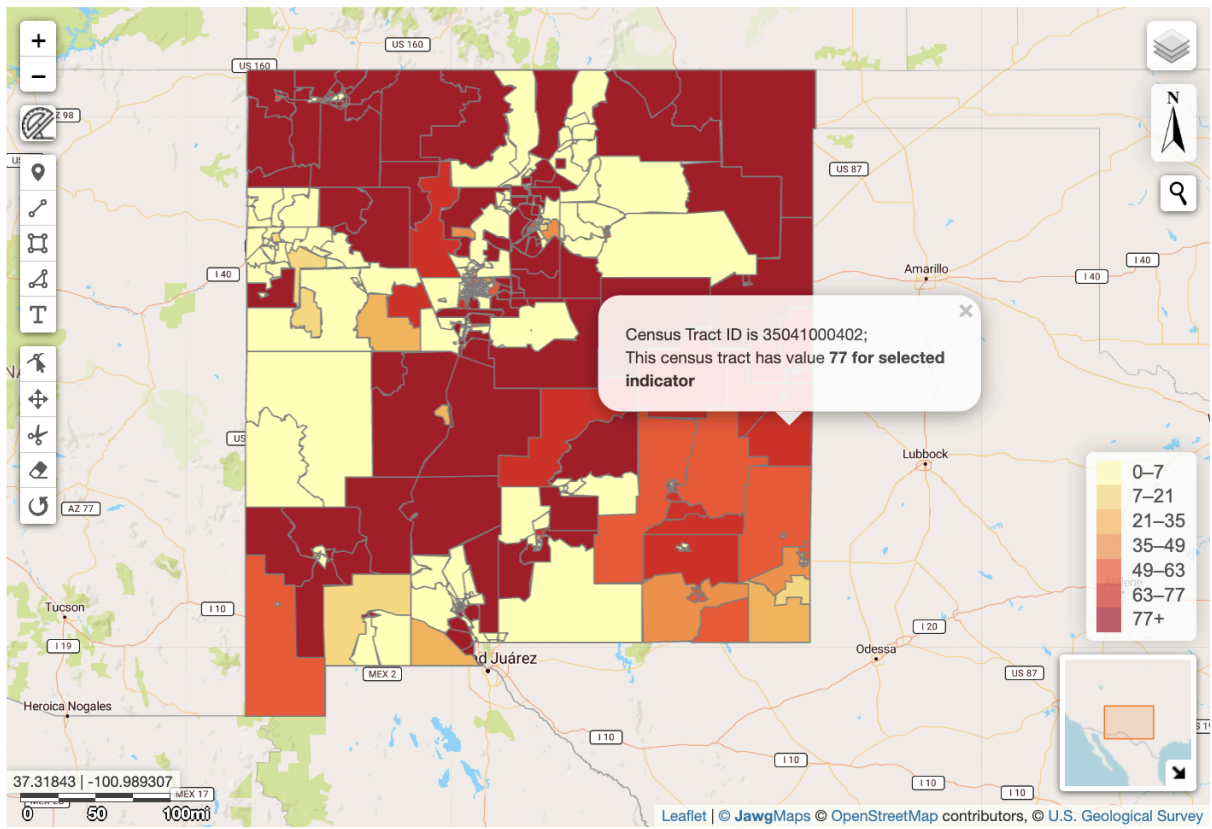

Figure 6. Choropleth map with dropdown for selecting climate and economic indicators in iNMsocialJusticeMap. Overview charts (Figures 9a, 10a) are placed at the top of the web app to support all three maps. The poverty-rate histogram (Figure 10b) and related donut charts (Figures

9b, 9c, 9f) are positioned beside the map to enable intuitive comparison and interaction with shared thematic data. Note: A brief demo video illustrating interactive features (Figures 6–12) is available on the iNMsocialJusticeMap page (see DATA AND CODE AVAILABILITY STATEMENT).

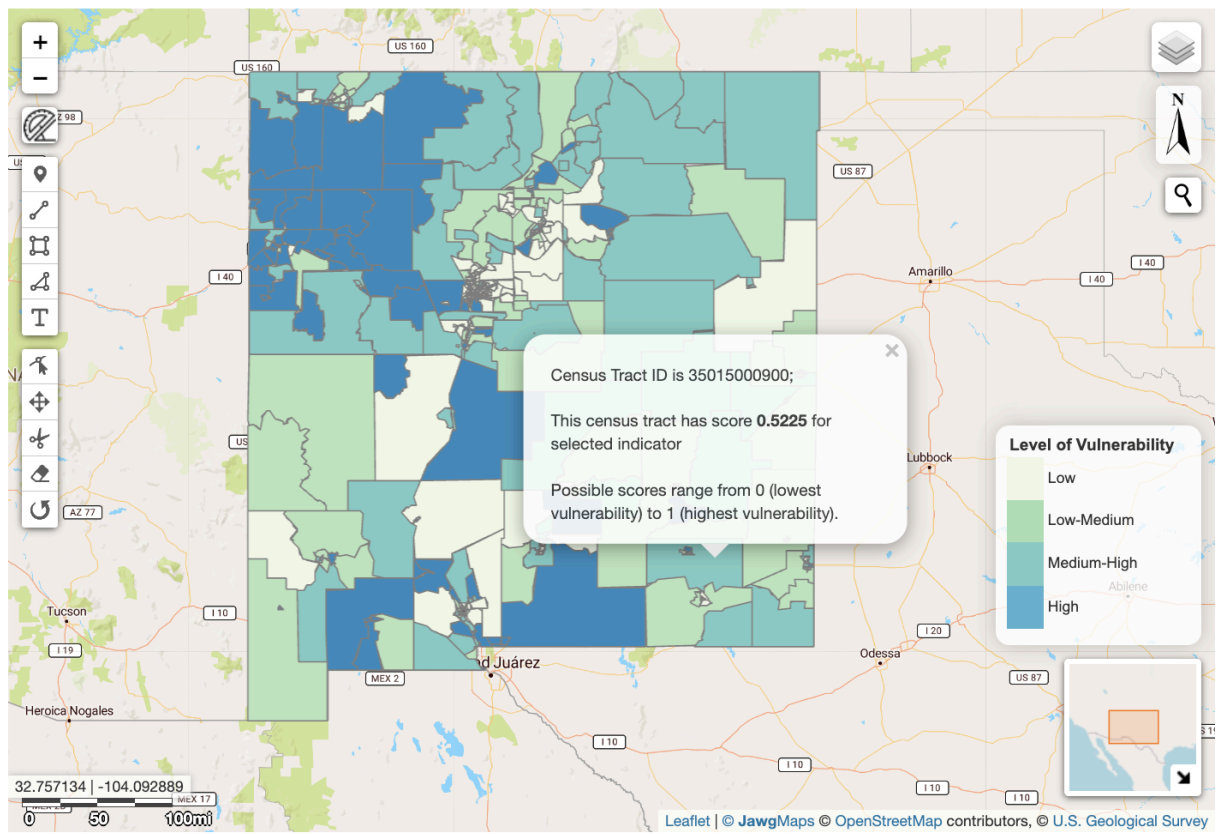

Figure 7. Choropleth map in iNMsocialJusticeMap displaying Social Vulnerability Index (SVI) indicators, selectable via a dropdown menu. Adjacent histograms (Figure 10c–f) show socioeconomic status (c), household characteristics (d), racial/ethnic minority status (e), and housing/transportation (f), enabling interactive exploration and thematic comparison.

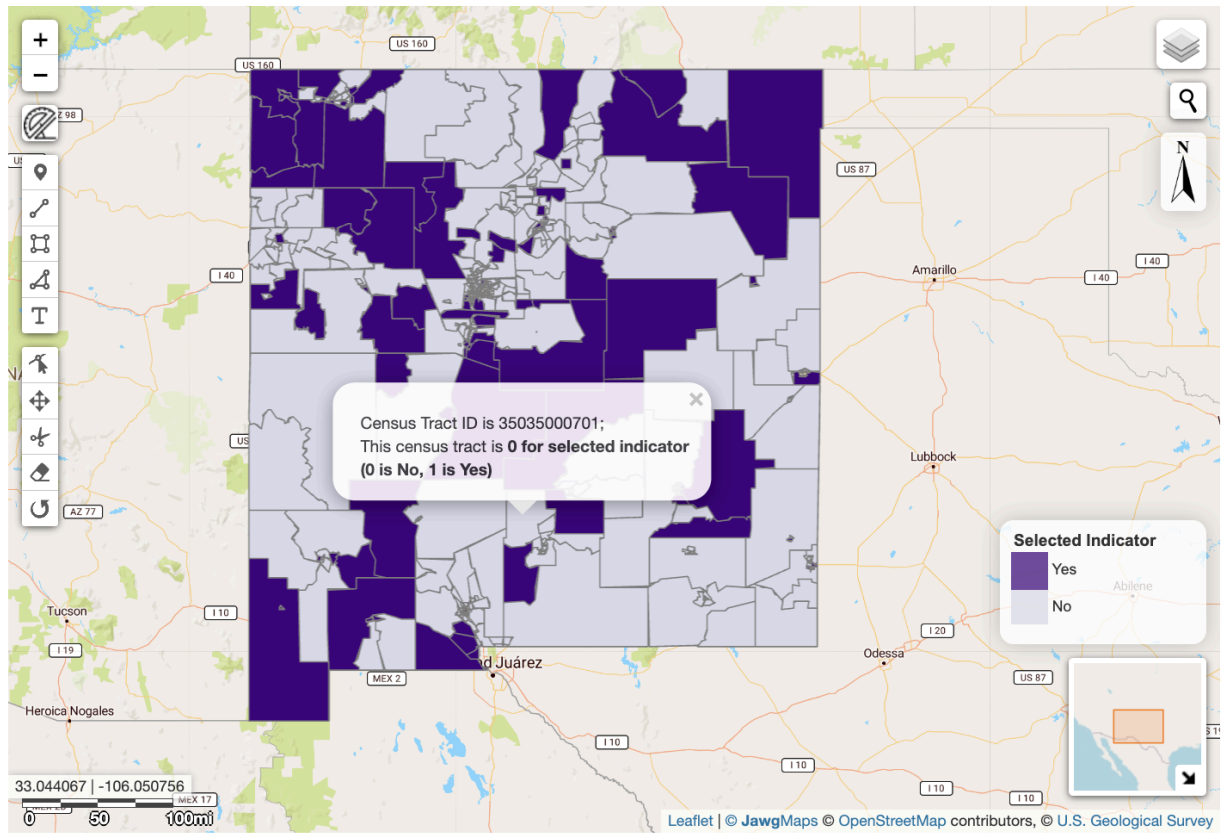

Figure 8. Choropleth map in iNMsocialJusticeMap visualizing food access, with a dropdown for selecting specific indicators. Adjacent donut charts show census tracts with low food access at 1 mile (Figure 9d) and 10 miles (Figure 9e). A bar chart displays Supplemental Nutrition Assistance Program (SNAP) participation by tract (Figure 9g), supporting interactive and thematically aligned exploration. Additional charts (not shown in this figure ) cover 20 miles and ½ mile without vehicle access. See the DATA AND CODE AVAILABILITY STATEMENT for access to not-shown charts and interactive versions of the charts and maps in all figures 3 to 15.

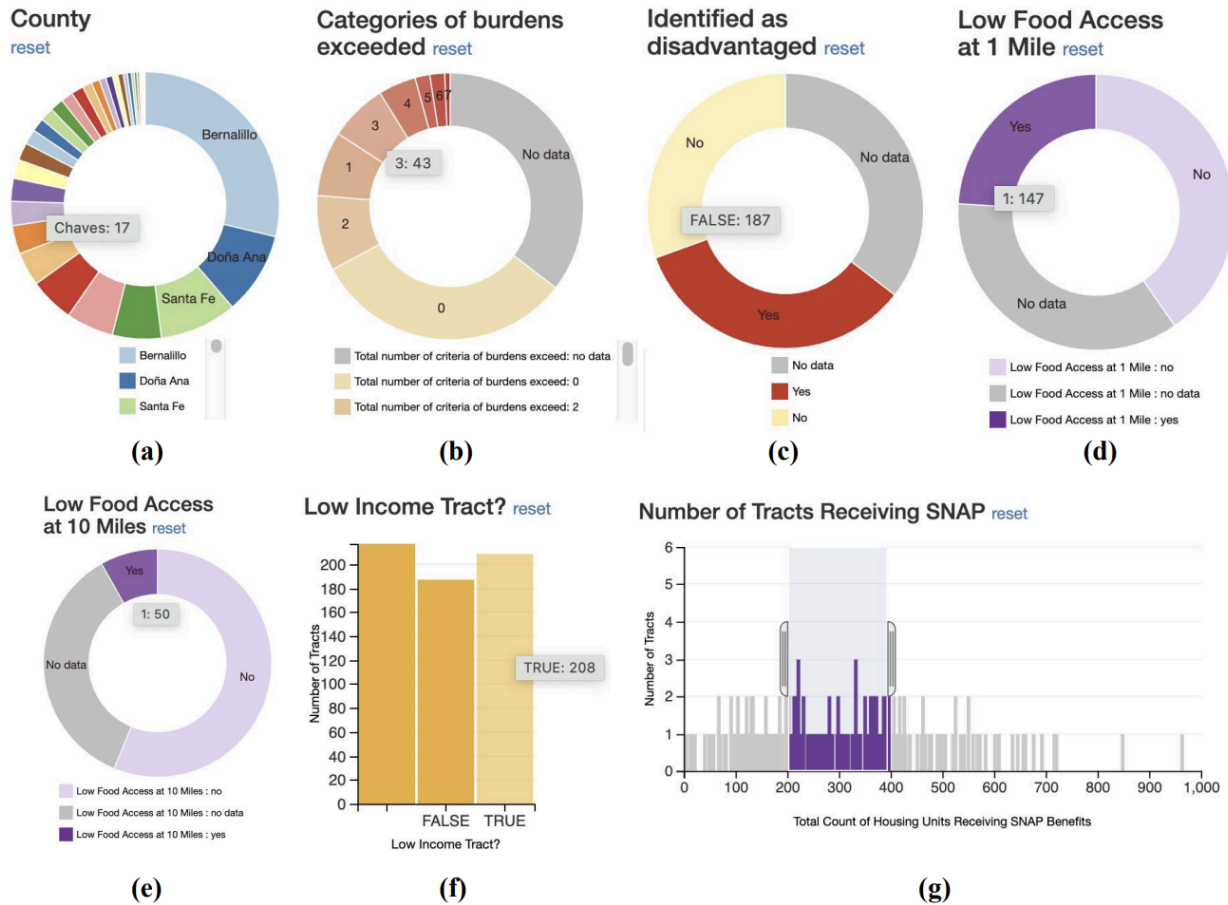

Figure 9. Donut charts with scrollable legends and bar charts in iNMsocialJusticeMap. Donut charts (a–e) display county distribution, burden categories, disadvantaged status, and low food access at 1- and 10-mile thresholds. *Scrollable legends* enhance readability and save app layout space. (f–g) Bar charts show low-income tracts and SNAP participation, with clicking (shown in f) and brushing functionality (illustrated in g) enabled for interactive filtering.

*dciWebMapper2*   by Sarigai Sarigai and Liping Yang *et al*.                                              25/46

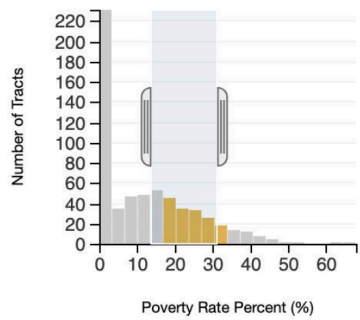
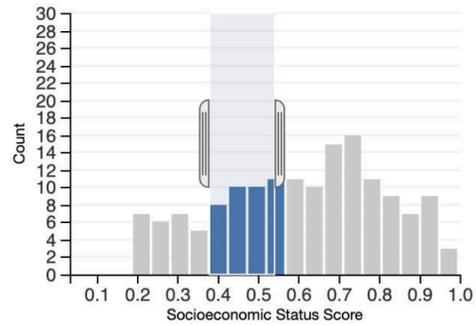
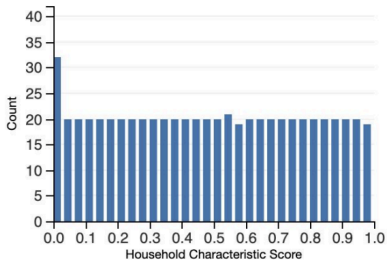
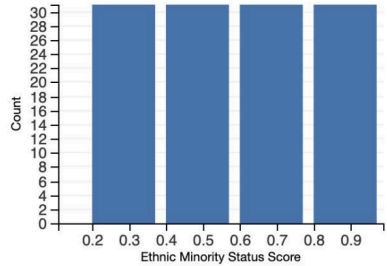
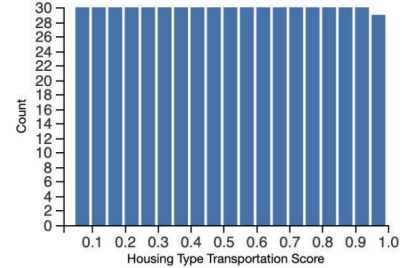

Figure 10. Interactive visualizations in iNMsocialJusticeMap. (a) *Select menu* showing census tract IDs (first three selected). Histograms (b–f): poverty rate (b), socioeconomic status (c), household characteristics (d), minority status (e), and housing/transportation (f), each with a dropdown-enabled bin size control. Different bin sizes are shown across (d–f) to illustrate flexibility in data exploration. Brushing is enabled in all histograms (b, c shows active brushing; b, d, e, f have brushing function available but not active in the figure) for dynamic filtering and exploration.

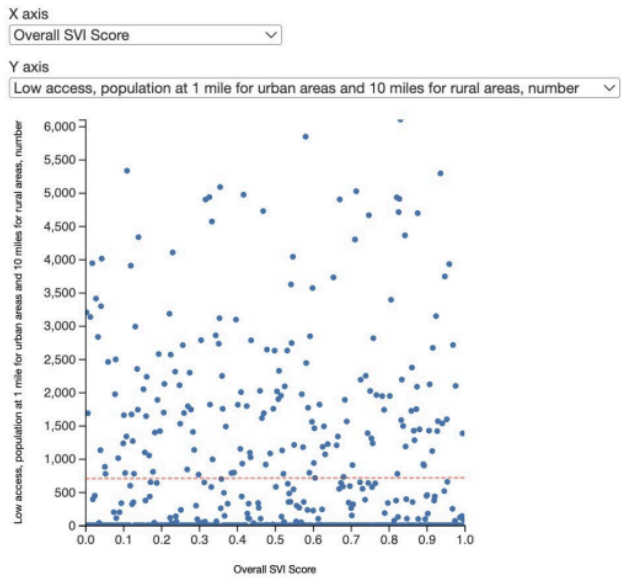
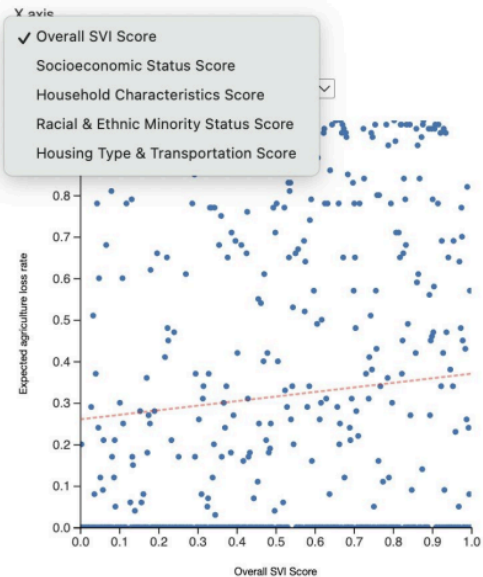
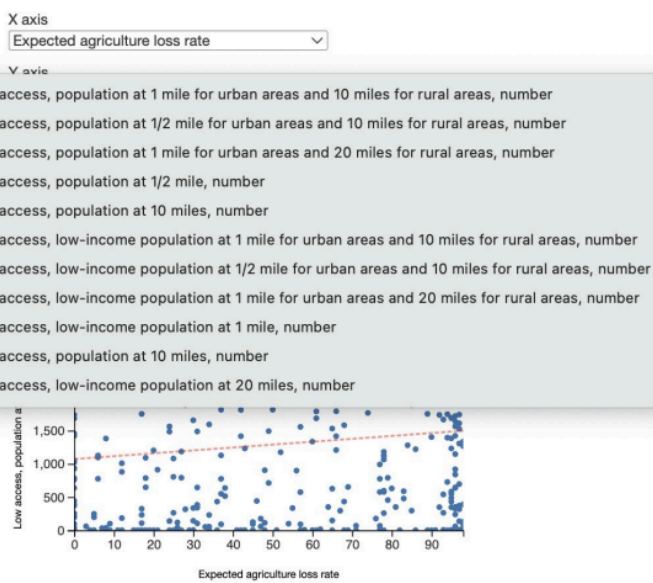
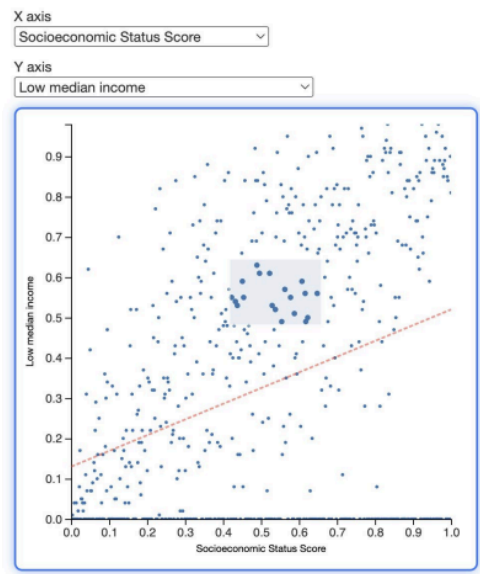

Figure 11. Interactive regression charts in iNMsocialJusticeMap. (a) Scatterplot with regression line linking SVI and Food Access indicators with axis selection. (b) X-axis dropdown for selecting SVI and Climate & Economic Justice indicators. (c) Y-axis dropdown for Climate & Economic Justice and Food Access indicators. (d) Scatterplot with regression line with polygon-based brushing to highlight user-selected data points for focused analysis. The dropdown-enabled interface empowers users to dynamically and intuitively explore variable relationships, revealing insights not easily accessible or even impossible in static charts. Brushing allows users to investigate subsets (e.g., outliers or points near the fitted line) and view their geospatial patterns and associated environmental and social vulnerability attributes.

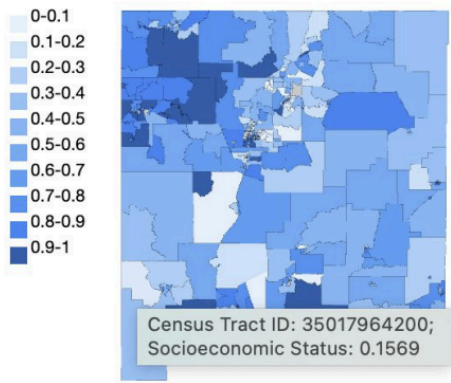
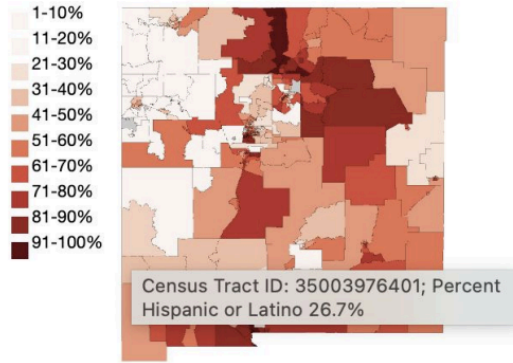

(a)            (b)

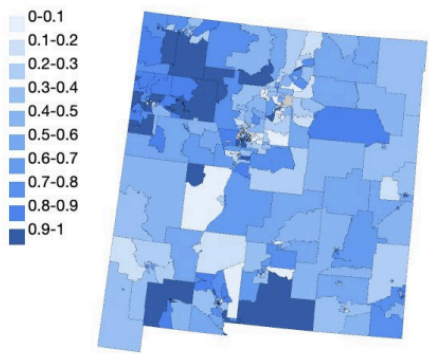
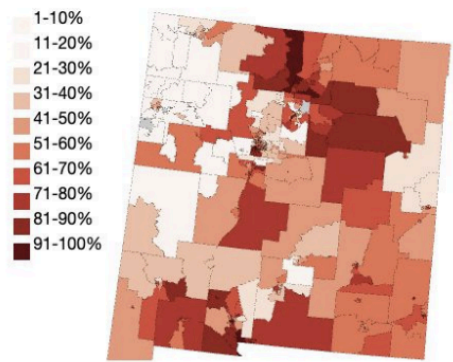

(c)            (d)

Figure 12. Small multiple maps illustrating demographic and socioeconomic variables in iNMsocialJusticeMap: (a) Socioeconomic status using the default projection, (b) Hispanic/Latino population distribution with the same projection as (a), and (c, d) Socioeconomic status and Hispanic/Latino population with Albers equal-area projection selected via dropdown. Small multiples support side-by-side geographic comparison, while the projection dropdown enables users to examine how map projections affect spatial perception.

## 4.4. iABQtrafficCrashMap web map app
4.4.1. Introduction, motivation, and data collection

Road traffic safety is a global concern, with 1.19 million deaths reported in 2021 and crash-related costs projected at $1.8 trillion from 2015–2030 (Amiri et al., 2021; World Health Organization, 2018). Crashes are the leading cause of death for individuals aged 5–29. In New Mexico, 2023 statistics show a person was injured every 28 minutes, killed every 20 hours, and a crash occurred every 34 minutes in Bernalillo County (UNM-GPS, 2023). Addressing this public health challenge requires data-driven approaches to understand crash patterns and inform prevention. Albuquerque's Vision Zero initiative and statewide efforts, such as the UNM-GPS crash database, reflect this commitment to improving roadway safety through targeted, evidence-based interventions (City of Albuquerque, 2023). Interactive mapping tools play a vital role by enhancing user engagement and supporting informed decision-making.

Our iABQtrafficCrashMap case study uses police-reported crash data from the New Mexico Department of Transportation, managed by UNM's Geospatial and Population Studies group (UNM-GPS, 2023). The database includes all public roadway crashes causing death, injury, or at least $500 in property damage, categorized as fatal, injury, or property damage-only. From 2010 to 2023, over 4,000 fatal and over 71, 000 injury crashes were recorded in Bernalillo County in New Mexico. These data were obtained by submitting a formal request to the University of New Mexico's Geospatial and Population Studies (GPS) Center via their online data‑request portal (https://gps.unm.edu/tru/request-data.html). For visualization, the original dataset was cleaned to exclude records lacking coordinates or containing georeferencing errors to ensure mapping clarity and accuracy.

4.4.2. Web app design and implementation

The implemented iABQtrafficCrashMap (https://geoair-lab.github.io/iABQtrafficCrashMap/index.html) provides an interactive platform to explore fatal and injury crash patterns in Albuquerque, New Mexico's highest-incident area, using police-reported crash data from 2010–2023 (geocoded by UNM-GPS). Focusing on severe crashes, the app enables users to analyze spatial and temporal trends based on attributes such as date, city, severity, and involvement of alcohol, drugs, pedestrians, bicycles, motorcycles, or tractors, supporting data-informed interventions aligned with Albuquerque's Vision Zero goals.

**Map visualizations (Figures 13 and 14).** Figure 13a shows county-wide crash distribution, with interactive pop-ups displaying detailed crash records upon user click. A built-in spatial filtering tool (Figures 13b–c) allows users to select geographic areas (e.g., a polygon, radius, and bounding box), to isolate crashes within specific zones. An instructional module (Figure 13b) guides the process, and filtered results are displayed dynamically on the map (Figure 13c). Additional layers including county boundaries, Native American lands, and crash heatmaps, are accessible via the layer control panel (click the top-right layer icon in Figure 13c to view layers shown in Figure 13d). **Two heatmap modes**, the last two layer options shown in Figure 13d, include: (1) a static/global heatmap view displaying overall crash density, and (2) a dynamic/local heatmap view that updates based on user filters and brushing (illustrated in Figure 14). A red-to-blue gradient highlights crash hotspots and low-density zones for rapid spatial pattern insight. See Figure 14 caption for why both views are necessary.

**Charts and interactive features.** Figure 15 presents interactive components adapted from both *dciWebMapper* and *dciWebMapper2* to support flexible filtering and analysis: (a) A *date slider* enables temporal filtering by year, month, and date. (b) *Select menus* allow filtering by city, county, severity, and other crash attributes. (c) *Row charts* display factors like alcohol, drug, and pedestrian involvement for deeper contextual insights.

Together, these interactive features offer an intuitive, multi-dimensional interface for exploring crash data, identifying high-risk areas, and supporting evidence-based decision-making for roadway safety.

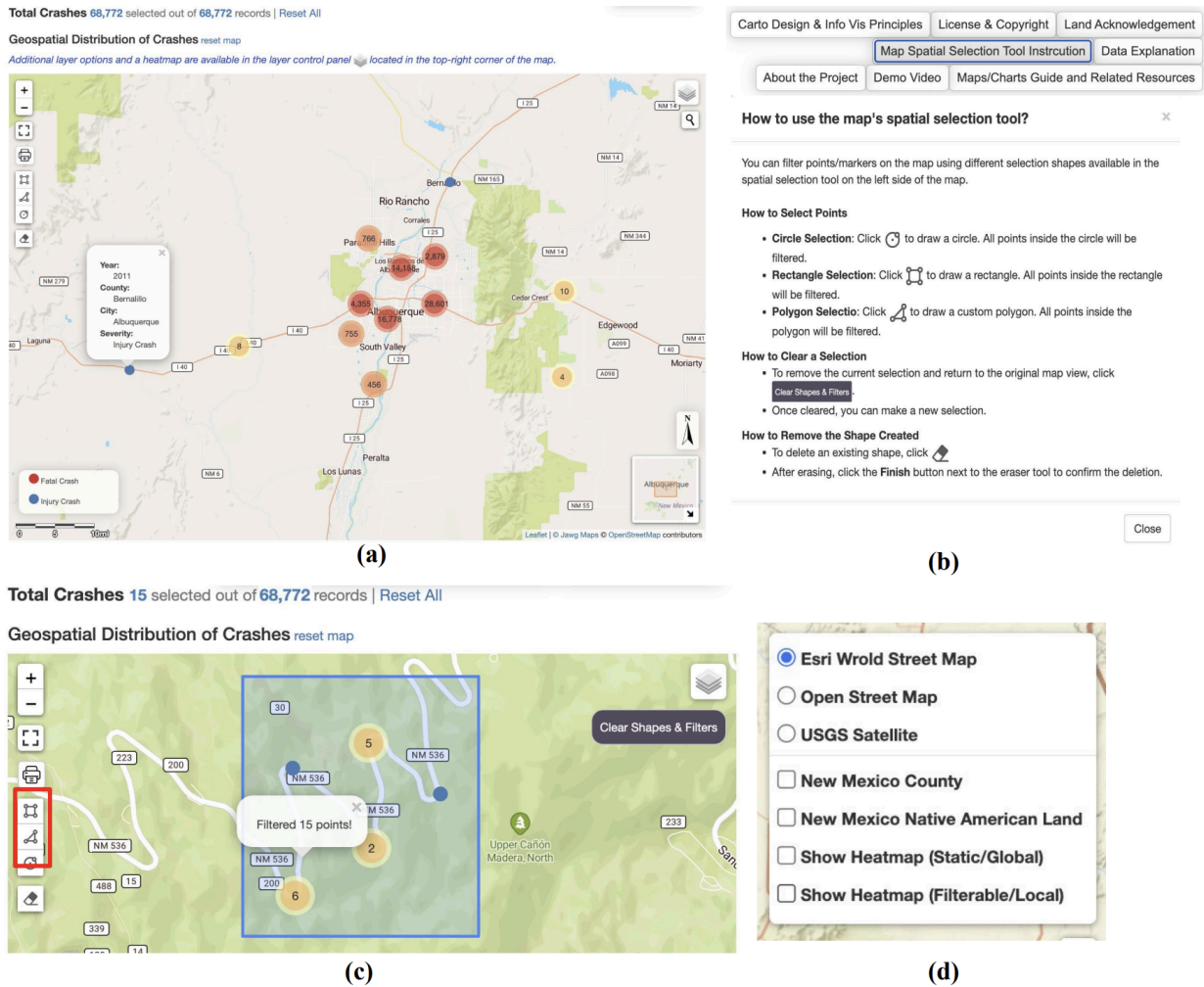

Figure 13. Geospatial distribution of the traffic crashes in iABQtrafficCrashMap. (a) Interactive map of reported crash locations. (b) Instructions for the spatial selection tool, accessed on iABQtrafficCrashMap via clicking the "Map Spatial Selection Tool Instruction" tab highlighted in Figure 13b.(c) Filtered map view based on a user-defined area (e.g., polygon, radius, or bounding box), following the instructional steps shown in (b). (d) Layer control panel showing additional overlays, county boundaries, Native American lands, and crash heatmaps, accessible via the top-right corner layer icon in (c). A brief demo video showcasing interactive features (Figures 13–15) is available on the iABQtrafficCrashMap page (see DATA AND CODE AVAILABILITY STATEMENT).

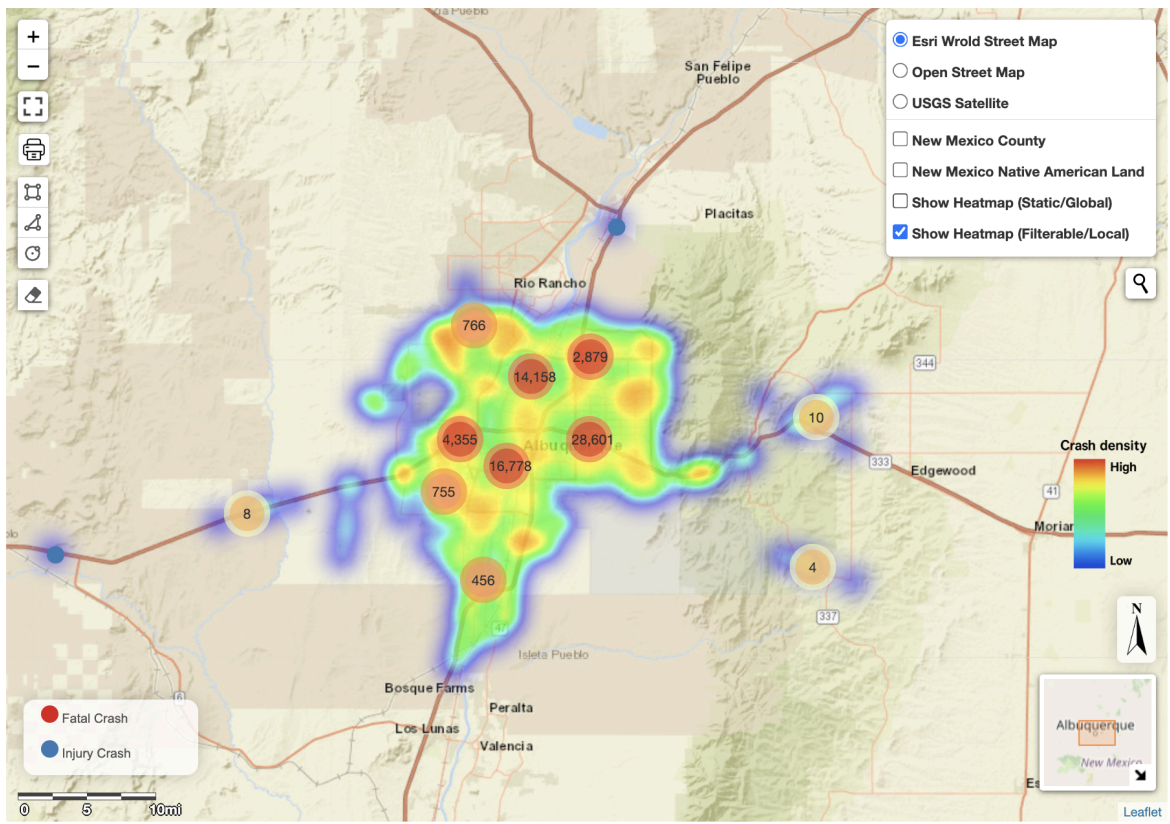

Figure 14. Traffic crash density visualized in the local heatmap, one of two heatmaps in iABQtrafficCrashMap. Enabled via the layer control panel (top-right corner), this raster-based heatmap shows crash density across Albuquerque, NM. Users can toggle between a static/global view of overall patterns and a filterable/local version that responds to interactive selections. Note: Because the cluster marker layer cannot be turned off, it may obscure parts of the heatmap, especially in high-crash areas, making it difficult to interpret the underlying density patterns from heatmap. In these cases, users can zoom into dense areas and switch to the global (static) heatmap view, where cluster markers are divided into smaller groups, which will reduce visual obstruction and improve pattern visibility from the global heatmap.

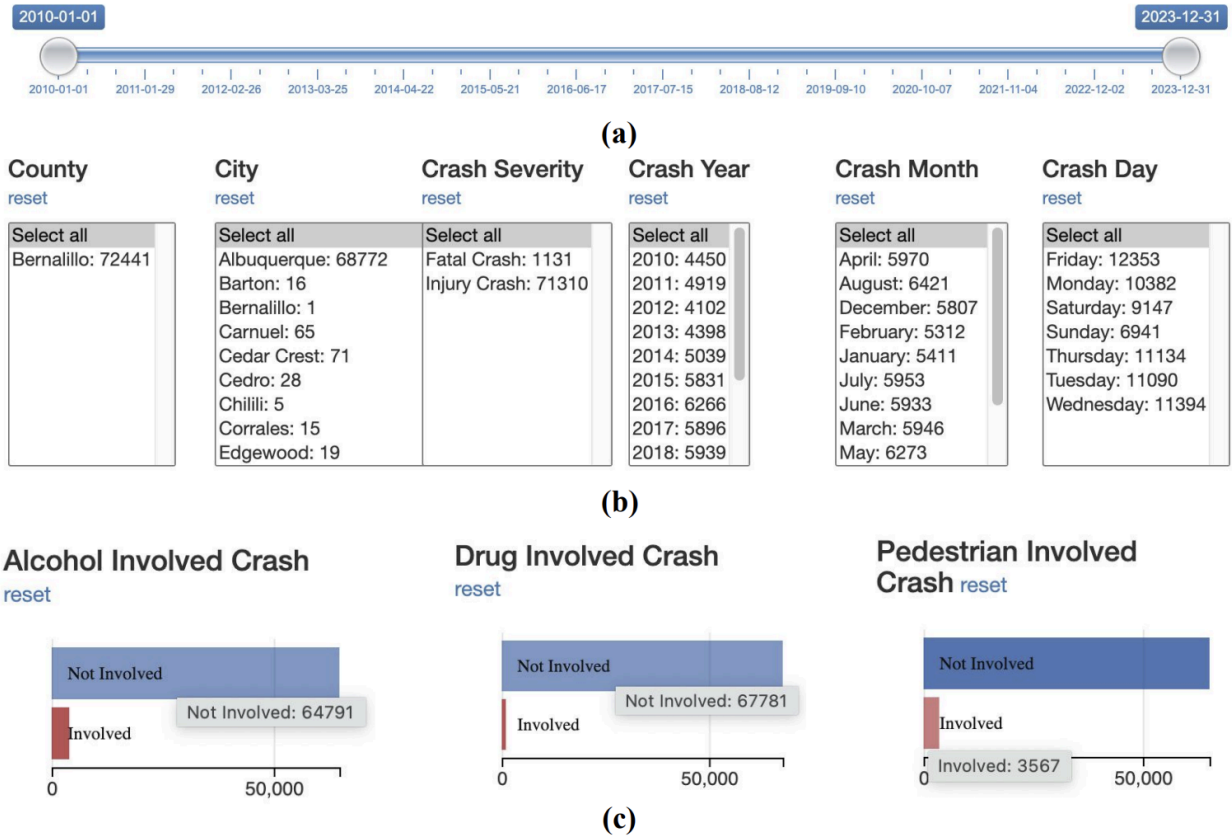

Figure 15. Interactive visualizations in iABQtrafficCrashMap include: (a) a date slider to filter crashes by year, month, and day; (b) select menus for county, city, severity, and time-related crash attributes; and (c) row charts showing involvement of alcohol, drugs, and pedestrians.

## 4.5 Reinforcing foundational design: Insights from case study applications of the extended framework

This section examines how the cartographic and information visualization principles established in the original *dciWebMapper* framework (Sarigai et al., 2025), particularly those in Section 4, are carried forward and applied in the expanded *dciWebMapper2* framework across three case studies. Rather than replacing the original, *dciWebMapper2* extends its modular, extensible design through technical and visual enhancements that reinforce the original design principles, demonstrating their adaptability across varied data types, interface components, and analytical contexts.

4.5.1. Reflections on information visualization principles in practice

This section offers reflections on how the expanded framework continues to embody the information visualization principles established in the original *dciWebMapper* framework (Sarigai et al., 2025). The following reflections correspond to the principles outlined in Sections 4.1.1- 4.1.5 of the original *dciWebMapper* framework paper (Sarigai et al., 2025) (can also be accessed from the Carto & Info Vis Design Principles" modules in the corresponding web map app (e.g. https://geoair-lab.github.io/iABQtrafficCrashMap/Carto&InfoVisDesignPrinciples.html), demonstrating how these foundational guidelines were upheld and extended across the expanded design and case study implementations.

1) To reduce visual clutter and enhance clarity in dense spatial datasets, *dciWebMapper2* incorporates techniques such as marker clustering and dropdown-driven layer selection. These design decisions are grounded in cognitive research showing that clutter negatively impacts visual search and user performance (Rosenholtz et al., 2007). More specifically, a) each *dciWebMapper2* web app includes only essential components: a primary web map, a semantically and interactively linked set of charts, complementary map types, and a data table. In iNMsocialJusticeMap (Section 4.3), three primary maps are paired with their own focused charts and elements, maintaining semantic coherence and visual clarity. b) Each layout follows a minimalist design, removing decorative or redundant elements while retaining purposeful interface redundancy to support comprehension (Tindall-Ford et al., 1997). For example, in iABQtrafficCrashMap (Section 4.4), the heatmap layer is optional via the control panel, allowing users to activate it as needed without overwhelming the default view. c) Marker cluster maps are used in dense point datasets to reduce clutter while preserving spatial context. d) In iNMsocialJusticeMap, dropdown-driven choropleth maps render one variable at a time, helping users focus on specific topics (e.g., climate burden or food access) without overloading the interface, also improving performance. e) Corresponding legends for dropdown-driven choropleth maps are dynamically generated based on the selected variable, further streamlining interpretation. Together, these five design strategies ensure that each component serves a clear, intentional role. The resulting interfaces are approachable, even for users with limited GIS experience, and support effective pattern recognition, comparison, and insight generation without distraction.

2) We iteratively refined the three case study web map apps developed with *dciWebMapper2* through multiple internal rounds of feedback and design reviews conducted within our research lab. While each app underwent continuous improvement during development, formal usability testing was only conducted on the most advanced and comprehensive iNMsocialJusticeMap app (see Section 4.3), which integrated multiple complex and closely related datasets (partial authors of this paper, in preparation). This app was selected as the representative evaluation case because it embodied nearly all of the expanded capabilities of the *dciWebMapper2* framework, including multivariate filtering, coordinated-view interactions across diverse charts and maps, small multiples, and dropdown-driven dynamical rending of choropleth maps and dropdown-driven map projection selection for map layers, except for the time slider, proportional symbol maps, and spatial filtering, which were demonstrated in the other two web map apps. Its integration of multiple thematically related datasets (spanning social, environmental, and food justice domains) made it especially well-suited for assessing usability, interpretability, and analytical utility. This evaluation thus offered meaningful insight into the *dciWebMapper2* framework's effectiveness while also indirectly validating its broader applicability for complex geovisualization tasks. Because the web map app addressed socially relevant topics and featured an intuitive interface, it was well-suited for participants across a broad range of expertise and mapping experience, making it an appropriate platform for assessing the usability and effectiveness of the expanded

*dciWebMapper2* framework. In contrast, the other two web map applications were either too domain-specific or narrowly focused in scope, making them less suitable for evaluating broad usability and framework adaptability across diverse user groups.

3) The selection of appropriate chart and map types for each of the three case study web apps was guided by the chart design recommendations outlined in Section 3.4 of the *dciWebMapper* framework paper (Sarigai et al., 2025). For the newly introduced visual components in the expanded framework *dciWebMapper2*, including additional chart and map types, further design guidance based on data type compatibility is provided in Section 3.5. Each of the three case study web apps illustrates how these recommendations were applied to support effective spatial data exploration, with detailed rationale discussed in Sections 4.2.2, 4.3.2, and 4.4.2 in the original *dciWebMapper* framework (Sarigai et al., 2025).

4) In terms of layout, *dciWebMapper2* introduces a flexible design that aligns with core information visualization principles, particularly the emphasis on highlighting key content while minimizing distractions. These new layout options enhance adaptability for organizing visual components based on dataset characteristics and user goals while maintaining visual clarity (Wesson, 2017). For example, the single-map layout (Figure 1), used in both the original *dciWebMapper* case studies and in the iPathogenTrackingMap (Section 4.2) and the iABQtrafficCrashMap map (Section 4.2), centers attention on a primary map interface. This clean, centralized structure supports multiple layers and interactions without overwhelming the user, embodying the principle of keeping the important elements prominent and simplifying surrounding visual context. In contrast, the row-based vertical multi-map layout (Figure 2) used in iNMsocialJusticeMap (Section 4.3) enables direct comparison of three distinct datasets, each presented in its own horizontal section with a dedicated map and corresponding charts. This vertically structured layout enhances readability, reduces scrolling, and supports clear spatial reasoning across thematic domains, particularly for state- and city-level analyses. The layout choice is guided by data characteristics, geographic scale, and visualization goals, ensuring usability and clarity. To preserve focus and minimize visual noise, supporting content such as explanatory text and metadata is placed on separate pages, in line with the principle of emphasizing essential content while simplifying or deferring secondary elements.

5) Reflecting the principle of using color wisely, the *dciWebMapper* framework prioritizes intentional and accessible color choices to support clear, coherent interpretations. All three case study applications in *dciWebMapper2* (Section 4.2 to Section 4.4) adopt color-blind-friendly palettes to enhance inclusivity and readability. In the iNMsocialJusticeMap app, consistent color schemes are applied across each of the three choropleth maps, their associated set of interactive charts, and the corresponding data table columns, reinforcing thematic cohesion, supporting more intuitive visual interpretation, and thus reducing cognitive load. Similarly, the iPathogenTrackingMap and iABQtrafficCrashMap maintained unified thematic colors across components, while selectively using broader color ranges in charts to differentiate numerous categories without overwhelming the viewer. This approach exemplifies a balance of accessibility, aesthetics, and interpretability, ensuring that color enhances rather than distracts from spatial insight.

4.5.2. Reflections in terms of Cartographic Principles

This section reflects on how the expanded framework reinforces the core cartographic principles established in *dciWebMapper* (Sarigai et al., 2025) through new case studies.

1) Map aesthetics, though inherently subjective, play a critical role in effective web map design. In *dciWebMapper2*, we emphasize visual coherence by harmonizing symbols, colors, and labels to create a consistent and intuitive user experience. In our iNMsocialJusticeMap, to intuitively distinguish justice domains, each domain uses a distinct dominant color theme: (1) the SVI map retains the CDC's familiar blue palette for continuity; (2) climate and economic justice indicators use a red-to-yellow gradient to signal environmental risk; and (3) the food access map adopts a purple scheme for visual contrast. These color themes extend to associated charts, reinforcing thematic clarity and supporting cross-comparison. In iNMsocialJusticeMap, unified palettes are applied across climate and economic justice (CEJST), SVI, and food-access visualizations. In iABQtrafficCrashMap, marker and chart colors consistently reflect crash severity. In iPathogenTrackingMap, distinct colors differentiate CHOV-positive and negative samples for quick interpretation. Across all three apps, *dciWebMapper2* balances visual appeal with clarity, enabling clear interpretation of complex spatial patterns.
2) The web map applications built with *dciWebMapper2* prioritize clarity, usability, and thoughtful design. Each layout includes core map elements (e.g., a scale bar and north arrow) within a clean, focused interface (Wesson, 2017). Interface redundancy is intentionally applied using coordinated symbols, colors, and text to support diverse cognitive processing preferences (Tindall-Ford et al., 1997). Concise titles help users immediately grasp map purpose, while preattentive visual cues (e.g., color, line width, text size) guide attention to key features. To reduce clutter, scalable marker clustering is used in iPathogenTrackingMap and iABQtrafficCrashMap, dynamically adjusting with zoom. All three web apps include compact base map and layer toggles to maintain visual balance (Wesson, 2017). Projection choice is also treated as a design element: while Web Mercator is used for tile-based maps, iNMsocialJusticeMap and iPathogenTrackingMap support projection switching (e.g., Albers Equal-Area, cylindrical, azimuthal) for improved spatial accuracy in choropleth maps.
3) Our *dciWebMapper2* incorporates four core cartographic design principles, adapted from Shneiderman (2003) and Roth & Harrower (2008), summarized by Janicki et al. (2016), and emphasized in Sarigai et al. (2025), to support effective, aesthetically engaging interactive web mapping. These principles are reflected across the three case study web apps. In iPathogenTrackingMap and iABQtrafficCrashMap, marker maps offer an overview by default, with details on demand via clickable markers and layer controls. In iNMsocialJusticeMap, dropdown menus allow users to toggle between justice-related indicators to dynamically update each of the three choropleth maps, enabling deeper, user-driven exploration. Across all three apps, geospatial data is presented through maps, charts, and tables, accommodating varied cognitive styles. Users can interact with the data through multiple pathways, such as dropdown selections, brushing histograms, and clicking map units or chart elements, reinforcing the principle of multiple perspectives and flexible interaction.
4) All three case studies in our *dciWebMapper2* use the Web Mercator projection for tile-based maps (i.e., the marker and proportional symbol maps in

iPathogenTrackingMap, the marker map in iABQtrafficCrashMap, and the main choropleth maps in iNMsocialJusticeMap). As the standard for web mapping, Web Mercator provides a familiar, scalable view, though it does not preserve area, an important limitation for choropleth maps. To address this, *dciWebMapper2* introduces projection-switching functionality for SVG-based maps, enabling users to choose among Albers equal-area conic, cylindrical, and azimuthal projections. This added flexibility supports more accurate spatial interpretation for area-based analyses and allows users to tailor visualizations to the geographic scope and analytical needs of their data.

**5. Discussion**

The enhanced framework, *dciWebMapper2,* advances interactive web cartography by integrating multi-type maps, statistical charts, and coordinated-view visualizations within an open-source, modular structure. Supporting point, choropleth, and proportional symbol maps with time-enabled sliders and interactive charts enable dynamic exploration of spatial, temporal, and thematic patterns. Through diverse use cases, including small multiples and dynamic choropleth comparisons, the framework fosters transparency, adaptability, reproducibility, and multivariate analysis, empowering users to explore complex geospatial dynamics across domains and time.

**5.1. Strengths**

The enhanced framework *dciWebMapper2* combines robust technical functionality with conceptual advances, supporting interactive web cartography and broader GIScience applications through integrated exploration of spatial, temporal, and attribute data (Sections 5.1.1–5.1.2).

5.1.1. Technical strengths and implementation features

The *dciWebMapper2* integrates a lightweight, modular architecture with advanced interactivity and analytical functionality. Its extensible codebase supports clarity, maintainability, and adaptability across domains. The framework enables coordinated views between maps, charts, and spatial tools, allowing rich exploration of spatial, temporal, and attribute patterns. Features include dropdown-based attribute switching, small multiples, proportional symbol maps, choropleth maps, time sliders, and support for both point and polygon geometries. With Turf.js and Leaflet-Geoman, it incorporates web-based spatial analysis (e.g., point-in-polygon), extending lightweight web mapping toward desktop GIS-level capabilities and scalability.

1) **Lightweight and modular architecture:** The *dciWebMapper2* framework is built on a lightweight, browser-based architecture that requires no proprietary software or complex server infrastructure, ensuring ease of deployment and accessibility. Its modular codebase is organized into clearly defined components, with well-commented functions and reusable scripts, allowing developers to easily adapt, extend, or substitute elements for different application needs. All required libraries and dependencies are bundled within each web map application itself, making the framework fully self-contained and functional without relying on third-party hosting services for critical libraries. This structure promotes rapid customization, ease of maintenance, and seamless integration with other open-source tools, making the framework highly adaptable for a variety of geovisualization goals and contexts.
2) **Well-documented and extensible codebase:** The framework's codebase is organized using clear modular structures and is thoroughly commented to support readability, learning, and reuse. Core functions, configuration parameters, and data processing routines are logically separated, making it easy for developers, educators, and researchers to locate, modify, or extend specific components. By following consistent coding patterns and naming conventions, the *dciWebMapper2* framework lowers the barrier for contribution and customization. This design empowers users to build on the core functionality and create tailored applications for a variety of spatial data visualization needs with minimal configuration, without requiring extensive prior experience in web development.

3) **Integrated linked views for multi-modal visualization:** The *dciWebMapper2* framework implements coordinated-view interactions to enable dynamic linking across multiple map types and diverse statistical charts, allowing users to explore spatial, temporal, and attribute-based relationships within a unified interface. (1) Linked filtering, highlighting, and cross-selection enhance the analytical depth of the tool, supporting multi-faceted spatial reasoning and comparative analysis across diverse datasets and thematic domains (in particular, see Section 4.3 for the complex datasets used in the iNMsocialJusticeMap). (2) To address the common challenge of overlapping spatial records, particularly in point-based datasets such as pathogen surveillance or incident reporting, the tool integrates Leaflet MarkerCluster's spiderfy functionality. The spiderfying feature reduces visual clutter by separating coincident points that share identical spatial coordinates, thereby improving interpretability and interaction. For example, in pathogen tracking datasets, it is common for multiple pathogens to be recorded at the same host location; spiderfying these overlapping points allows users to more intuitively and clearly distinguish individual records (see Figure 3a, Section 4.2). (3) *dciWebMapper2* introduces interactive scatter plots with regression lines and customizable histograms, allowing users to explore correlations, distributions, and trends among continuous variables. These features expand the analytical depth of the platform beyond spatial comparisons, supporting more comprehensive data interpretation across multiple dimensions. (4) The framework also integrates spatial analysis tools via Turf.js and the Leaflet-Geoman plugin, enabling web-based geoprocessing functions such as point-in-polygon queries and spatial filtering. These analytical capabilities, traditionally available only in desktop GIS platforms, are now accessible within an interactive web environment, expanding the framework's utility for exploratory spatial analysis and decision support.

4) **Dropdown-based attribute switching and thematic mapping:** The *dciWebMapper2* framework enables dynamic exploration of multivariate data through intuitive dropdown-driven controls that allow users to switch between various mapped attributes or indicators. This functionality supports flexible thematic exploration through comparative analysis across variables while reducing visual clutter by avoiding multiple overlapping map layers. Users can seamlessly toggle between metrics related to food access, climate vulnerability, social conditions, and other domain-specific indicators, facilitating high-dimensional data interpretation without sacrificing usability or performance. These advanced layout and filtering controls also accommodate both exploratory (high interaction) and quick-reference (low interaction) use cases, making the interface adaptable to diverse analytical needs across decision-making, education, and public communication contexts.

5) **Support for high-dimensional, data-driven exploration:** The *dciWebMapper2* framework enables intuitive and dynamic visualization and comparison of complex, multivariate geospatial data through interactive features such as dropdown-driven choropleth maps, small multiples, proportional symbol maps, and time sliders. These expanded map types and capabilities allow users to explore spatial, temporal, and attribute-based patterns in an integrated environment, supporting dynamic comparisons and nuanced analysis of complex geospatial datasets. Additionally, these expanded map types offer flexible ways to visualize and compare spatial phenomena across points and polygons, variables, and timeframes, supporting exploratory analysis of both attribute-rich and geometry-diverse datasets over time, all within a unified, interactive environment.
6) **Scalability across geographic and temporal scales:** The framework is designed to handle spatial data at various scales, from census tracts to regional or national levels. It accommodates both fine-resolution and aggregated geographies, allowing users to conduct analyses that range from local patterns to broader spatial comparisons. Additionally, temporal interaction tools, such as time sliders, enable the exploration of changes over time, supporting longitudinal studies and dynamic data storytelling. This scalability ensures the framework's applicability across a broad range of use cases, including localized mapping, longitudinal studies, and comparative spatial analysis across different regions or periods.

5.1.2. Impactful conceptual strengths

The conceptual strengths of *dciWebMapper2* stem from both its technical architecture and broader geovisualization impact. Its clarity, reusability, transferability, and reproducibility support open, modular design adaptable across disciplines. As an open-source, extensible tool, it empowers exploratory spatial analysis, decision-making, and inclusive engagement. Together, these qualities promote transparency, sustainability, and accessibility, enabling expert users, educators, students, and communities to interact with spatial data in open, meaningful, and user-centered ways.

1) **Clarity and reusability:** The *dciWebMapper2* framework is built with modularly structured functions, clear and extensive in-code documentation (i.e., commented code) and intuitive configuration, which enables users to quickly understand and adapt its components. This clarity enhances reusability and adaptability, which allows researchers, practitioners, and educators to extract and repurpose individual core modules (e.g., interactive charts, time sliders, map layers, dropdowns) across domains such as environmental monitoring, public health, and urban planning without steep learning curves or redundant development. This supports efficient development and fosters experimentation without requiring users to build tools from scratch. The framework's modular and transparent structure supports reuse in a wide range of spatial data applications beyond its initial context.
2) **Transferability:** The *dciWebMapper2* framework is designed to be data-agnostic, transferable, and adaptable across a wide range of thematic areas, including transportation safety, environmental and economic justice, and food security, by supporting diverse spatial data types and flexible visual configurations. Its generalizable structure, clear documentation, and modular architecture make it easy for users to apply the framework to different types of geospatial data and audiences. This transferability extends beyond technical reuse, supporting flexible integration across disciplines, use cases, and societal challenges.

3) **Reproducibility and transparency:** The *dciWebMapper2* framework is fully self-contained and openly shared, with detailed documentation, well-commented code, and example datasets, along with design principles aligned with open science practices, *emphasizing transparency, reproducibility, and accessibility.* The transparency allows others to understand the implementation logic, replicate the same visualization outcomes using the provided data and code, and adapt the interactive mapping framework with confidence. Together, these attributes support rigorous, reproducible workflows and align with open science practices in spatial research and communication.

4) **Open-source and extensible framework:** The *dciWebMapper2* framework is openly available and built with extensibility in mind, encouraging adaptation, customization, and continued development by a broader community. This openness not only supports transparency and collaboration but also ensures the framework can evolve alongside emerging needs in spatial analysis, cartographic storytelling, and interactive mapping tools. The framework enables users to *build custom geovisualization applications without reliance on proprietary tools or complex technical infrastructure.*

5) **Support for exploratory spatial analysis and decision-making:** *The three use case web map apps have demonstrated dciWebMapper2's ability to support decision-making and hypothesis generation by empowering users to explore spatial relationships and patterns across variables, scales, and time.*

6) **Accessibility and user empowerment:** The framework is intentionally designed with accessibility in mind, minimizing technical barriers for users with varying levels of coding and GIS/mapping expertise. Through intuitive configuration, clear documentation, and lightweight deployment, the applications developed through applying *dciWebMapper2* framework enable a broader range of users, including educators, students, and community stakeholders, to meaningfully interact with spatial data, explore patterns, and generate insights. In particular, those applications supporting independent exploration give users control to filter, compare, and analyze spatial and statistical data interactively, rather than just viewing static maps.

## 5.2. Limitations and future work

Despite its flexibility and enhanced functionality, *dciWebMapper2* has several limitations that suggest promising directions for future development. First, its fully client-side architecture, though self-contained, portable, and easy to maintain, can face performance bottlenecks when handling large or frequently updated datasets. Incorporating optional server-side components for spatial indexing or preprocessing could improve scalability while preserving its lightweight design. Second, the framework currently supports coordinated bivariate views but lacks multivariate, 3D, and more advanced thematic mapping features. Promising enhancements include adding bivariate and multivariate choropleth mapping and a 3D map layer option to deepen analytical capabilities and support multidimensional insight. Third, while *dciWebMapper2* includes tools like heatmaps and spatial queries, its analytical depth remains primarily visual. Expanding support for spatial statistics (e.g., spatial autocorrelation) could increase its value for exploratory spatial analysis. Collectively, these enhancements will advance

*dciWebMapper2* as a modular, open-source platform for interpretable, interactive, and analytically rich web cartography across diverse research and application domains.

## 6. Conclusion

This work presents the significant advancement of an open-source web mapping framework named *dciWebMapper* designed to meet the growing demand for flexible, transparent, and extensible geovisualization tools. By incorporating additional map types, temporal interactions, spatial analysis capabilities, and linked statistical charts, the enhanced framework *dciWebMapper2 empowers users to explore high-dimensional geospatial data across space, time, and thematic domains.* These capabilities are delivered through a fully self-contained, modular architecture that promotes reproducibility, long-term sustainability, and broad adaptability across disciplinary and applied contexts.

      Grounded in cartographic and information visualization design principles, the framework advances the practice of interactive web mapping by lowering technical barriers and promoting visual clarity, interpretability, and user engagement. Its application to domains such as climate and economic justice, food access, and social vulnerability, highlights its cross-sector relevance and capacity to support accessible spatial storytelling and informed decision-making. The open-source, modular nature of the framework makes it a valuable resource for researchers, educators, practitioners, and community organizations seeking to communicate and explore spatial data without relying on proprietary platforms. As interactive mapping becomes increasingly central to spatial analysis, public engagement, and digital scholarship, frameworks like this one help shape the evolving landscape of open, transparent, and data-rich geovisualization. This work lays the foundation for future extensions of the *dciWebMapper* and/or *dciWebMaper2* framework, including support for multivariate map visualizations and enhanced interoperability, aimed at enabling broader applications in spatial analysis and evidence-based decision-making. By leveraging open-source tools and prioritizing modular design, the *dciWebMapper2* framework is well-positioned to evolve alongside emerging geospatial data needs and analytical paradigms. As geospatial storytelling and spatial data science continue to converge, this framework has the potential to become a versatile resource for researchers, practitioners, and educators seeking to create transparent, interactive, and reproducible web-based geovisualizations, advancing big data-driven insights and supporting informed, evidence-based decisions.

## Endnotes

1. https://github.com/IonDen/ion.rangeSlider
2. https://d3js.org/
3. https://square.github.io/crossfilter/
4. https://dc-js.github.io/dc.js/
5. https://leafletjs.com/
6. https://datatables.net/
7. https://jquery.com/
8. https://getbootstrap.com/
9. https://turfjs.org/
10. https://github.com/geoman-io/leaflet-geoman
11. https://msb.unm.edu/
12. https://stridata-si.opendata.arcgis.com/datasets/SI::corregimientos-census-2010-feature-layer/about

**Acknowledgments**

This work was partially supported by the National Science Foundation (NSF) award (NSF 2155222) and by the funding support at the University of New Mexico from the College of Arts and Sciences, and from the Office of the Vice President for Research WeR1 Faculty Success Program (WeR1 FaST 2022 and WeR1 SuRF 2022). The authors are also grateful for the valuable discussions and insights contributed by collaborators on the NSF project, particularly the following individuals: Drs. Jocelyn P. Colella and Marlon Cobos (Biodiversity Institute and Department of Ecology and Evolutionary Biology, University of Kansas, Lawrence). Dr. Folashade Agusto (Department of Ecology and Evolutionary Biology, University of Kansas, Lawrence), Dr. Emma Goldberg (Theoretical Biology & Biophysics at Los Alamos National Laboratory and Los Alamos New Mexico Consortium); the authors would like to particularly thank Dr. Jonathan L. Dunnum (Division of Mammals, Museum of Southwestern Biology at the University of New Mexico) and Dr. Blas Armién (Instituto Conmemorativo Gorgas de Estudios de la Salud in Panama) for providing the host sample dataset visualized in the iPathogenTrackingMap. The authors would also like to thank all eight authors from the *dciWebMapper* paper published in *Transactions in GIS* (Sarigai et al. 2025) and two students (C.J. Allen and Jessica Hilfers) who took GEOG 485L in Spring 2023 at the University of New Mexico. The authors also thank the editors and anonymous reviewers for their valuable feedback and constructive suggestions.

**Disclosure statement**
No potential conflict of interest was reported by the author(s).

**Data availability statement**
Interactive versions of all maps and charts presented in Figures 3–15, including additional not-shown visualizations, are accessible via the associated *dciWebMapper2* case study web apps. The three case study web map apps accompanying this paper are all open-source and can be found at the following links:
- *iPathogenTrackingMap*: https://geoair-lab.github.io/iPathogenTrackingMap/index.html (**demo video link:** https://www.youtube.com/watch?v=mLvbDq-xzVM).
- *iNMsocialJusticeMap*: https://geoair-lab.github.io/iNMsocialJusticeMap/index.html (**demo video link:** https://www.youtube.com/watch?v=ilrbP3PvYF8).
- *iABQtrafficCrashMap*: https://geoair-lab.github.io/iABQtrafficCrashMap/ (**demo video link**: https://www.youtube.com/watch?v=kX9CpD3q3jg)